\documentclass[aps,twocolumn,english,superscriptaddress,citeautoscript,showkeys,preprintnumbers,amsmath,amssymb,floatfix,footinbib,prb] {revtex4-2}
\usepackage{graphicx}
\usepackage{xcolor}
\usepackage{soul}
\usepackage{amssymb,amsmath}
\usepackage{hyperref}
\hypersetup{colorlinks=true, linkcolor=red, citecolor=blue}
\usepackage[utf8]{inputenc}
\usepackage[english]{babel}
\usepackage{txfonts}
\usepackage[font=normalsize, caption=false]{subfig}

\newcommand{\tc}{$T_\text{c}$}
\newcommand{\mustar}{$\mu^{*}$}
\newcommand{\nef}{$N_{\varepsilon_\text{F}}$}
\newcommand{\omegalog}{$\omega_{\log}$}
\newcommand{\af}{$\alpha^2F$}

\begin{document}


\title{Ab initio modeling of superconducting alloys}

\author{P. N. Ferreira}
\email[Corresponding author: ]{pedroferreira@usp.br}
\affiliation{Computational Materials Science Group (ComputEEL/MatSci), Universidade de S\~ao Paulo, Escola de Engenharia de Lorena, DEMAR, Lorena, Brazil}
\affiliation{Institute of Theoretical and Computational Physics, Graz University of Technology, NAWI Graz, 8010, Graz, Austria}
\author{R. Lucrezi}
\affiliation{Institute of Theoretical and Computational Physics, Graz University of Technology, NAWI Graz, 8010, Graz, Austria}
\author{I. Guilhon}
\affiliation{Grupo de Materiais Semicondutores e Nanotecnologia (GMSN), Instituto Tecnológico de
Aeronáutica (ITA), 12228-900 S\~ao José dos Campos/SP, Brazil}
\author{M. Marques}
\affiliation{Grupo de Materiais Semicondutores e Nanotecnologia (GMSN), Instituto Tecnológico de
Aeronáutica (ITA), 12228-900 S\~ao José dos Campos/SP, Brazil}
\author{L. K. Teles}
\affiliation{Grupo de Materiais Semicondutores e Nanotecnologia (GMSN), Instituto Tecnológico de
Aeronáutica (ITA), 12228-900 S\~ao José dos Campos/SP, Brazil}
\author{C. Heil}
\email[Corresponding author: ]{christoph.heil@tugraz.at}
\affiliation{Institute of Theoretical and Computational Physics, Graz University of Technology, NAWI Graz, 8010, Graz, Austria}
\author{L. T. F. Eleno}
\email[Corresponding author: ]{luizeleno@usp.br}
\affiliation{Computational Materials Science Group (ComputEEL/MatSci), Universidade de S\~ao Paulo, Escola de Engenharia de Lorena, DEMAR, Lorena, Brazil}

\begin{abstract}

Designing new, technologically relevant superconductors has long been at the forefront of solid-state physics and chemistry research. However, developing efficient approaches for modeling the thermodynamics of superconducting alloys while accurately evaluating their physical properties has proven to be a very challenging task. To fill this gap, we propose an ab initio thermodynamic statistical method, the Extended Generalized Quasichemical Approximation (EGQCA), to describe off-stoichiometric superconductors. Within EGQCA, one can predict any computationally accessible property of the alloy, such as the critical temperature in superconductors and the electron-phonon coupling parameter, as a function of composition and crystal growth conditions by computing the cluster occurrence probabilities that minimize the overall mixing Gibbs free energy. Importantly, EGQCA incorporates directly chemical ordering, lattice distortions, and vibrational contributions. As a proof of concept, we applied EGQCA to the well-known Al-doped MgB$_2$ and to niobium alloyed with titanium and vanadium, showing a remarkable agreement with the experimental data. Additionally, we model the near-room temperature sodalite-like Y$_{1-x}$Ca$_x$H$_6$ superconducting solid solution, demonstrating that EGQCA particularly possesses a promising potential for designing \emph{in silico} high-\tc{} superhydride alloys. Our approach notably enables the high-throughput screening of complex superconducting solid solutions, intrinsically providing valuable insights into the interplay between synthesis, thermodynamics, and physical properties. 
   
\end{abstract}

\date{\today}


\maketitle

\section{Introduction}
 
Properties of materials are primarily engineered by alloying and introducing defects and impurities. This is particularly true in the case of superconductors, where the incorporation of dopants, even in small concentrations, can transform ordinary metals and semiconductors into superconductors \cite{morosan2006,wang2008,brauner2009,wray2010,hor2010,renosto2013,machado2017,barbero2017,de2018,wu2020,bhattacharyya2022,correa2022,correa2023}. 

The increasingly \emph{in silico} design of novel high-temperature, low-pressure superconducting materials highlights the importance of doping and the controlled inclusion of crystalline defects~\cite{di2021, lucrezi2022, wang2022, saha2020, di2022, ferreira2023}. These methods are crucial for fine-tuning the electronic structure, phonon dispersion, and electron-phonon coupling of these systems at or close to  ambient conditions. Consequently, developing efficient first-principles approaches for modeling superconducting alloys is highly desirable and would significantly impact the advancement of technologically relevant superconductors. 

\begin{figure*}[t]
	\includegraphics[width=.7\linewidth]{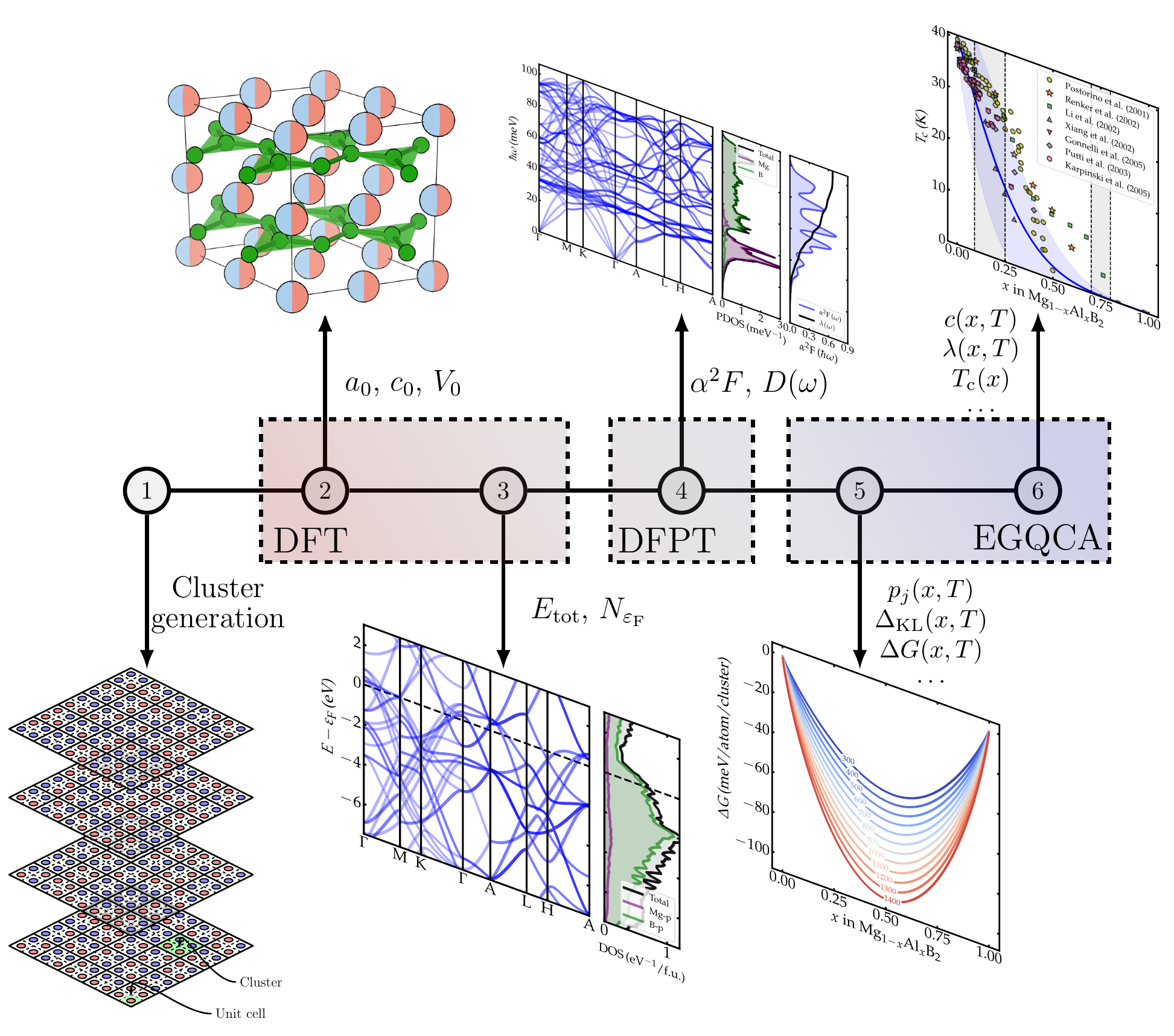}
	\caption{The EGQCA workflow encompasses the following steps: (1) Construction of non-equivalent clusters using symmetry space group analysis; (2) Structural optimization and (3) electronic structure calculations employing the DFT formalism; (4) Calculation of the phonon frequencies and electron-phonon coupling through the first-order linear response theory within DFPT; (5) Application of the EGQCA minimization to calculate thermodynamic quantities and (6) the determination of any computationally accessible physical property as a function of temperature and composition utilizing the cluster occurrence probability that minimizes the global Gibbs mixing free energy. This approach allows for comprehensive and accurate modeling of alloyed superconductors, enabling the prediction of their physical properties and behavior.}
	\label{fig:abstract}
\end{figure*}

There are essentially two different approaches to treating alloyed systems from first principles. First, in non-structural methods, substitutionally random alloys are described in terms of symmetry-preserving perturbations of the end-member constituents by considering the average occupation of lattice sites, as seen in the virtual crystal approximation (VCA) \cite{dargam1997, bellaiche2000}. While this approach is useful for many applications~\cite{ramer2000, ma2005, heil2015}, it neglects local correlations, structural relaxation, and short- or long-range ordering. On the other hand, structural approaches, such as the supercell method, can explicitly incorporate the microscopic atomic structure. However, even in the simplest case of $n$ permutable sites in a crystal lattice composed of two elements, there are $2^n$ possible structural configurations, making density functional calculations impractical without further approximations. To address this challenge, the construction of periodic special quasirandom structures (SQS) \cite{zunger1990, wei1990} -- mimicking the correlation functions of an infinite substitutional random alloy by selectively distributing the atomic species among the disordered sites -- has become popular in the past few years. This approach has emerged as the standard procedure for calculating optical and electronic properties of disordered compounds, especially semiconductor alloys~\cite{zunger2003, van2004, freysoldt2014}, but it is not an appropriate choice for modeling the full solubility range of superconducting alloys as it requires the construction of computationally demanding large cells for electron-phonon calculations for each target composition. Besides, even though SQS can describe atomic disorder rather accurately at 0\,K, it is not a thermodynamic model \emph{per se},  in the sense that temperature effects and variations in composition are not necessarily taken into account unless SQS calculations are used as input data in an explicit thermodynamic model \cite{Zhu2021}.

Here, instead, we propose an alternative supercell approach to treat alloyed superconducting materials using fully ab initio data in a thermodynamic model, adapting an established theory in semiconductor physics \cite{chen1995} known as the Generalized Quasi-Chemical Approximation (GQCA) \cite{sher1987}, that resembles in some level effective cluster expansion methods such as the Cluster Variation Method (CVM) \cite{Kikuchi1951}. Within GQCA, the alloy is described as an ensemble of small, non-equivalent ordered supercells, each statistically and energetically independent of its local atomic environment. The occurrence probability of each supercell at a given composition and temperature is calculated analytically by minimizing the mixing Gibbs free energy, incorporating both the mixing enthalpy and configurational entropy. By computing the total energy and the physical properties of interest for each supercell, this method allows us to describe binodal and spinodal phase decompositions, miscibility gaps, order-disorder tendencies, and the evolution of any computationally accessible physical property, such as the electron-phonon coupling ($\lambda$) and superconducting critical temperature ({\tc}), as a function of composition and crystal growth temperature across the entire compositional range. 

We believe this methodology offers significant practical advantages over SQS or other related methods for studying superconductivity, in the sense that: $(i)$ with just a few non-equivalent, small supercells, we can describe all compositions of the solid solution, whereas SQS requires a unique structural configuration for each composition; and $(ii)$ it allows us to work with disordered systems while preserving exact, rather than effective, local interactions. Moreover, we extend this model by incorporating the vibrational free energy through the calculation of the phonon density of states within density functional perturbation theory (DFPT), resulting in an improved thermodynamic description beyond the original formulation of GQCA. Due to the new implementations, we hereby propose to call our method Extended Generalized Quasi-Chemical Approximation (EGQCA).

As a proof of concept, we apply the EGQCA to the Al-doped MgB$_2$ and niobium alloyed with titanium and vanadium. These systems have been extensively investigated theoretically and experimentally in the past and they exhibit a regular solid solution with complex features, making them perfect systems to benchmark our model. To demonstrate the predictive power of EGQCA, we also modeled, for the first time, the near-room temperature Y$_{1-x}$Ca$_x$H$_6$ superconducting solid solution, showcasing EGQCA as a particularly useful methodology for designing superhydride alloys. 

This paper is structured as follows: Section \ref{sec:gqca} presents a detailed formulation of the EGQCA and we discuss its usage in particular cases, specifically when electron-electron interactions significantly impact the material's superconducting properties and when some clusters are dynamically unstable. In Section \ref{sec:results}, we apply the EGQCA to four different systems: aluminum-doped magnesium diboride (Section \ref{sec:mgalb2}); niobium alloyed with titanium and vanadium (Section \ref{sec:nb}); and sodalite-like yttrium-calcium hydride (Section \ref{sec:ycah6}). Each case highlights a specific aspect of EGQCA. In Section \ref{sec:mgalb2}, we demonstrate how to interpret the derived cluster fraction probabilities and thermodynamic potentials, and how to use them to elucidate experimental results, providing new physical insights to Mg$_{1-x}$Al$_x$B$_2$. Section \ref{sec:nb} discusses how to address dynamically unstable clusters and end-members, how to accurately account for electron-electron interactions in calculating the \tc{} of conventional superconductors using EGQCA, and how to determine the spinodal and binodal curves of the solid solution, demonstrating the importance of the mixing vibrational free energy to correctly predict the miscibility gap for Nb$_{1-x}$V$_{x}$. Finally, in Section \ref{sec:ycah6}, we predict a novel high-\tc{} superhydride solid solution and showcase the promising potential of EGQCA for designing new superconducting alloys towards ambient conditions. We conclude with a summary of our findings in Section \ref{sec:summary} and the methodology in Section \ref{sec:methods}.

\section{Extended Generalized Quasichemical Approximation}
\label{sec:gqca}

\textbf{The GQCA Model.} In GQCA \cite{sher1987}, an alloy -- often but not necessarily a solid solution -- is represented by an ensemble of \(M\) individual clusters. These clusters are statistically and energetically independent of the surrounding atomic configuration and are spatially homogeneous on a macroscopic length scale. 

The original formulation of the GQCA model is thoroughly detailed in Refs.~\cite{teles2000, guilhon2015, guilhon2017}. Traditionally, GQCA is expressed in terms of the Helmholtz free energy, which is equivalent to the Gibbs free energy at zero pressure. Below, we will introduce the main aspects of the GQCA formulation using the Gibbs free energy, incorporating the pressure energy term. Yet, the GQCA equations can be readily extended to include ternary and quaternary alloys \cite{marques2003}. For simplicity, however, we will focus on the binary and pseudobinary descriptions.

In a binary solution with formula A$_{1-x}$B$_x$, atomic species A and B are supposed to substitute one another at one or more crystallographic sites. The clusters are arranged into $J$ non-equivalent classes, defined by their atomic distribution, each cluster with a distinct total energy $E_{j}$ and degeneracy $g_{j}$, indexed by $j = 1,2,\ldots,J$. Within this cluster ensemble, the occurrence probability of each cluster $j$ at a given composition and temperature is determined by the fraction of clusters, $p_{j}$, where
\begin{equation}
p_{j} = \frac{M_{j}}{M}.
\end{equation}
Here, $M_{j}$ represents the number of same-class clusters with energy $E_{j}$ and $M = \sum_{j = 1}^{J}M_{j}$ is the total number of clusters. In the following, we will refer to this quantity as cluster occurrence probability, or simply cluster probability.

From the classical definition of the Gibbs free energy, one can define the system's Gibbs mixing free energy as
\begin{align}
    \Delta G = \Delta H - T\Delta S,
    \label{eq:Delta-F-GQCA}
\end{align}
where $\Delta H$ is the mixing enthalpy and $\Delta S$ is the mixing entropy. Within harmonic approximation, $\Delta H$ is expressed as
\begin{align}
\Delta H(x, T, P) = M\sum_{j=1}^{J} p_{j} \Delta_{j},
\label{eq:delta_U_2}
\end{align}
where the reduced excess enthalpies, $\Delta_{j}$, are defined by
\begin{align}
\Delta_{j} = H_{j} - \dfrac{n - n_{j}}{n}H_{\text{A}} - \dfrac{n_{j}}{n}H_{\text{B}}, 
\end{align}
with $H_{j}$ being $H_{j} = E_{j} + PV_{j}$, where $P$ is the pressure of the system and $V_{j}$ is the unit cell volume of cluster $j$ at a given temperature and pressure. $H_{\text{A}}$ and $H_{\text{B}}$ are the enthalpies of the reference states for the pure elements (corresponding to the cluster fully occupied by either A or B atoms), $n$ the total number of sites in the clusters, and $n_{j}$ the number of atoms of type B in the cluster $j$. It is important to note that we consider the excess energy per permutable site, whereas the original GQCA formulation uses excess energy per supercell.

The mixing entropy is obtained from the classical Boltzmann definition within a microcanonical ensemble:
\begin{align}
	\Delta S(x, T, P) = k_\text{B} \ln W,
	\label{eq:delta_S}
\end{align}     
where
\begin{align}
	W = \dfrac{N!}{N_{\text{A}}!N_{\text{B}}!}\dfrac{M!}{\prod_{j=1}^{J}M_{j}!}\prod_{j=1}^{J}{\left(p_{j}^{0}\right)}^{M_{j}}
	\label{eq:W}
\end{align}
is the total number of possible configurations of a random, disordered solution formed by the cluster configurations $M_1, M_2, \ldots, M_{j}$; $N$ is the number of atoms in the crystal; $N_{\text{A}}$ and $N_{\text{B}}$ are the total numbers of atoms of type A and B (with $N=N_{\text{A}} + N_{\text{B}}$); $k_\text{B}$ is Boltzmann constant; and	
\begin{equation}
    p_{j}^0 = g_{j} x^{n_{j}}(1-x)^{n-n_{j}}
\end{equation}
is the cluster probability for a random distribution of A and B atoms at a given composition $x$ in the cluster with degeneracy $g_{j}$. In Eq.~(\ref{eq:W}), the first factor [$N!/(N_{\text{A}}! N_{\text{B}}!)$] is the number of complexions for a random binary distribution, whereas the remaining terms provide the multinomial probability that this random distribution will generate a configuration corresponding to each cluster. Eq.~(\ref{eq:W}) is therefore a generalization of the quasi-chemical approximation \cite{Howard1940Feb}, in which the clusters are simply nearest-neighbor pairs of atoms; hence the name of the method.

Since the clusters, albeit small in size, are spread throughout the whole crystal, and therefore $M$ is very large, we use Stirling's approximation $\ln \theta! = \theta \ln \theta - \theta + \mathcal{O}(\ln \theta$), so that the Gibbs mixing free energy within the original GQCA formulation is written as
\begin{align}
\Delta G(x,T, P) =& M\sum_{j=1}^{J} p_{j} \Delta_{j} + \nonumber\\ &+ Nk_{\text{B}}T\left[x\ln x + (1-x)\ln(1-x)\right] + \nonumber\\ &+ Mk_{\text{B}}T\sum_{j=1}^{J}p_{j}\ln\left(\dfrac{p_{j}}{p_{j}^0}\right).
\end{align} 
Finally, as the equilibrium state of a system maintained at constant temperature, pressure, and composition corresponds to the state of minimum Gibbs free energy, the unknown cluster probabilities $p_{j}$ will be those that minimize $\Delta G$ per cluster.

\textbf{Inclusion of vibrational effects.} Here, we propose the inclusion of the mixing vibrational free energy, $\Delta A$, into the GQCA formulation, so that we may rewrite Eq.~(\ref{eq:Delta-F-GQCA}) as
\begin{align}
	\Delta G(x,T, P) = \Delta H(x,T, P) - T\Delta S(x,T, P) + \Delta A(x,T),
\end{align}
where $\Delta A$ is given by
\begin{align}
	\Delta A(x,T) = M\sum_{j=1}^{J} p_{j} \varepsilon_{j}(T),
\label{eq:delta_A_2}
\end{align}
with
\begin{align}
\varepsilon_{j}(T) = A_{j}(T) - \dfrac{n - n_{j}}{n}A_{\text{A}}(T) - \dfrac{n_{j}}{n}A_{\text{B}}(T), 
\end{align}
in which $A_{\text{A}}(T)$, $A_{\text{B}}(B)$, and $A_{j}(T)$, the first two for the pure elements and the latter for cluster $j$, are obtained from the phonon density of states $D(\omega)$ using the harmonic approximation \cite{korotaev2018}:
\begin{align}
	A_{j}(T) = k_{\text{B}}T\int_{0}^{\infty} D_{j}(\omega)\ln\left[2\sinh\left(\dfrac{\hbar\omega}{2k_{\text{B}}T}\right)\right]\mathrm{d}\omega.
\end{align}
It should be noted that, as usual, thermal expansion effects are not included in the harmonic approximation. Therefore, as implemented, the EGQCA does not take into account the alloy volume as a function of temperature. We plan to implement the quasi-harmonic approximation \cite{Dove1993} in future versions of the method.

For the minimization of the free energy, Lagrange multipliers with respect to the constraints $\sum_{j=0}^{J} p_{j} = 1$ and $\sum_{j=0}^{J} n_{j}p_{j} = n x$ are introduced, resulting in cluster occurrence probabilities given by
\begin{align}
	p_{j} = \dfrac{g_{j}\eta^{n_{j}} e^{-(\varepsilon_{j} + \Delta_{j})/k_{\text{B}}T}}{\sum_{j=1}^{J}g_{j}\eta^{n_{j}} e^{-(\varepsilon_{j} + \Delta_{j})/k_{\text{B}}T}},
	\label{eq:xj}
\end{align}
where 
\begin{align}
	\eta = \dfrac{xe^{\lambda_{\text{L}}/k_{\text{B}}T}}{1-x},
\end{align}
and $\lambda_{\text{L}}$ are the Lagrange multipliers to be determined.

\textbf{Cluster occurrence probability.} For obtaining the $p_{j}\left(x,T\right)$ cluster probabilities, one has to solve the following $n$-order polynomial equation resulting from the minimization of the free energy:
\begin{align}
	\sum_{j=1}^{J}\left(nx - n_{j}\right)g_{j}e^{-(\varepsilon_{j} + \Delta_{j})/k_{\text{B}}T}\eta^{n_{j}} = 0,
\end{align}
which has a unique real solution.

The degree of similarity between the (E)GQCA cluster occurrence probability, $p_{j}$, and the probability distribution of the cluster $j$ in an ideal solid solution, $p_{j}^0$, can be quantified by the Kullback-Leibler divergence \cite{kullback1951}
\begin{align}
	\Delta_{\text{KL}}(x, T, P) = \sum_{j=1}^{J}p_{j}\ln\left(\dfrac{p_{j}}{p_{j}^0}\right). 
\end{align}
Complete randomness is achieved when $\Delta_{\text{KL}}(x,T)$ approaches zero, indicating that the temperature-dependent fractions $p_{j}$ equals the fractions $p_{j}^0$ of the regular solid solution model.

\textbf{Composition-dependent properties.} Within (E)GQCA, any computationally accessible, composition-dependent material property, $\mathcal{P}(x,T)$, can be determined as an average of the $\mathcal{P}_{j}$ properties corresponding to each cluster class $j$. These properties are calculated from fully relaxed supercells using ab initio methods (or, less commonly, from experimental measurements) and are weighted by the probabilities $p_{j}$ that minimize the overall Gibbs mixing free energy.:
\begin{align}
\mathcal{P}(x, T, P) = \sum_{j=1}^{J}p_{j}(x,T, P) \, \mathcal{P}_{j} \,.
\label{eq:EGQCA-props-avg}
\end{align} 
This procedure also enables evaluating the uncertainty of the theoretical estimation through the standard deviation associated with the distribution of the weighted $\mathcal{P}_{j}$ values:  
\begin{align}
\Delta \mathcal{P}(x,T, P) = \sqrt{ \sum_{j=1}^{J}p_{j}(x,T, P) \, \mathcal{P}_{j}^2 - \left[\mathcal{P}(x,T, P)\right]^2}.
\label{eq:EGQCA-props-std}
\end{align}
Our approach thus provides a robust method for comparison with experimental data, which is particularly crucial in cases where entropy significantly influences the results.

\textbf{Superconducting properties.} Regarding the description of the superconducting properties within (E)GQCA, one can obtain the Eliashberg spectral function for each non-equivalent cluster from DFPT \cite{DFPT} through the following expression 
\begin{align}
    \alpha^2F(\omega) = \dfrac{1}{2\pi N_{\varepsilon_{\text{F}}}}\sum_{\mathbf{q}\nu}\delta(\omega-\omega_{\mathbf{q}\nu})\dfrac{\gamma_{\mathbf{q}\nu}}{\hbar\omega_{\mathbf{q}\nu}},
\end{align}
where \nef{} is the density of states at the Fermi level and $\gamma_{\mathbf{q}\nu}$ is the phonon linewidth for mode $\nu$ at wavevector $\mathbf{q}$. With that one can compute the electron-phonon mass enhancement parameter,
\begin{align}
    \lambda = 2\int d\omega \dfrac{\alpha^2F(\omega)}{\omega}.
\end{align}
and estimate the critical temperature {\tc} for phonon-mediated superconductors using the Allen-Dynes formula \cite{allen1975}
\begin{align}
    \label{eq:AD}
    T_{\text{c}} &= \dfrac{f_{1}f_{2}\omega_{\text{log}}}{1.20}\exp\left(-\dfrac{1.04(1+\lambda)}{\lambda - \mu^{*}(1 + 0.62\lambda)}\right),\\
    f_1 &= \left(1 + \left(\dfrac{\lambda}{2.46(1 + 3.8\mu^{*})}\right)^{3/2}\right)^{1/3}, \\
    f_2 &= \left(1 + \dfrac{\lambda^2(\overline{\omega}_{2}/\omega_{\text{log}} - 1)}{\lambda^2 + \left[1.82(1 + 6.3\mu^{*})(\overline{\omega}_{2}/\omega_{\text{log}})\right]^2}\right),
\end{align}
where $f_1$ and $f_2$ are factors depending on $\lambda$, $\mu^{*}$, $\overline{\omega}_{2}$, and $\omega_{\log}$. Here, $\mu^{*}$ is the screened Couloumb pseudopotential according to the Morel-Anderson model~\cite{morel1962}, $\overline{\omega}_{2}$ denotes the 2nd root of the 2nd moment of the normalized distribution $g(\omega) = 2/\lambda\omega\alpha^2F(\omega)$, and $\omega_{\log}$ stands for the logarithmically average phonon frequency, given by
\begin{align}
    \omega_{\log} = \exp\left[\dfrac{2}{\lambda}\int\dfrac{d\omega}{\omega}\alpha^2F(\omega)\log(\omega)\right].
\end{align}

\textbf{(E)GQCA in practice.} At this point, we want to provide a practical guide on how to apply (E)GQCA to alloyed superconductors. Specifically, we want to discuss how to properly deal with the Coulomb pseudopotential while computing the \tc{} for the different clusters, and how to manage those systems that contain dynamically unstable configurations.   

\textbf{Electron-electron interactions.} As previously introduced, in the calculation of critical temperature for conventional superconductors based on Allen-Dynes formalism, Coulomb interactions are simplified to a single, adjustable semi-empirical parameter \mustar{}, introducing a certain degree of arbitrariness in \tc{} predictions. This issue becomes even more pronounced in alloys, as we are no longer dealing with a single, stoichiometric phase but rather the full range of solubility of a binary or ternary phase. It is, in principle, possible to obtain \mustar{} from \textit{GW} calculations, as for example demonstrated in Refs.~\cite{margine2016, heil2017, lian2019} , but doing this for every non-equivalent cluster of the solid solution is computationally prohibitive. Therefore, we must find reliable approximations describing how \mustar{} varies with composition ---  a challenging task, as discussed further in this work.   

Based on our expertise with EGQCA, we have established a standardized workflow for handling \mustar{} values in conventional superconductors within the Morel-Anderson approximation~\cite{morel1962}: $(i)$ if there are reliable experimental \tc{} data available for the end members, use them to derive the \mustar{} values for the terminal compounds then linearly interpolate \mustar{} for the disordered clusters. In this way, the number of adjustable parameters is kept to the strict minimum while considerably saving computational resources; or $(ii)$ if there is no experimental data and if computationally feasible, compute the \mustar{} values for the end members from first principles and linearly interpolate \mustar{} for the disordered clusters.

This procedure should provide an accurate description of the \tc{} behavior in alloys when the correlation and electron-electron renormalization effects are minor. In more complicated cases where spin fluctuations play a significant role, such as in Nb$_{1-x}$V$_x$ solid solutions~\cite{rapp1974}, the linear approximation of \mustar{} may break down, as we will demonstrate later. However, approximations $(i)$ and $(ii)$ should provide similar outcomes in cases where electron-electron interactions have a minor impact.

It is important to state that a more detailed yet computationally prohibitive approach would include computing the dynamical Coulomb interactions within Eliashberg formalism for each cluster. A first step into this direction is the inclusion of the static Coulomb interactions in Eliashberg theory from first principles, as described in Ref.~\cite{pellegrini2022}. Thus, while the calculation of \textit{W} is currently very expensive, we are confident that it will be possible in the future to avoid using any adjustable parameters in the calculation of superconducting \tc{} within the EGQCA, paving the way for fully \emph{in silico} high-throughput design of superconducting alloys.

\textbf{Dynamical stability}. The application of EGQCA to superconducting materials invariably necessitates the calculation of phonon spectra for each non-equivalent cluster. It is naturally expected that some atomic configurations may exhibit dynamic instability in the harmonic approximation, resulting in imaginary frequencies. Even when incorporating anharmonic contributions using advanced methods such as SSCHA~\cite{monacelli2021} or the special displacement method~\cite{zacharias2023}, some imaginary phonon modes may persist. Here, we argue that, even for systems where some clusters are not dynamically stable, (E)GQCA remains a robust approximation for modeling the behavior of solid solutions for the following reason.

By definition, a regular solid solution entails a random arrangement of solutes adhering to the crystal symmetry of the parent phase. Excluding unstable phases would introduce order into the solid solution, leading to an incorrect estimation of the mixing entropy, which predominantly influences the Gibbs mixing free energy at thermodynamic equilibrium.

One can think that, in real solid solutions, cluster-cluster interactions are presumed to renormalize the phonon dispersion, thereby locally stabilizing dynamically unstable atomic configurations. Given the impossibility of treating cluster interactions within (E)GQCA, integrating solely over stable, positive phonon modes to calculate the mixing vibrational free energy and the phonon-derived physical properties is a reasonable approximation as long as the majority of the alloy's phonon DOS is not discarded during the (E)GQCA averaging. For a practical example, see the case of Nb--Ti solid solution discussed in Section~\ref{sec:nb}.

\section{Results}
\label{sec:results}

\subsection{The case of Al-doped magnesium diboride}
\label{sec:mgalb2}

\textbf{The Al$_x$Mg$_{1-x}$B$_2$ solid solution.} MgB$_2$ is one of the most intriguing conventional superconductors. The hexagonal compound, characterized by a honeycomb boron lattice and a mixture of face and edge-sharing MgB$_{12}$ cuboctahedra, exhibits a {\tc} of 39\,K \cite{nagamatsu2001} and harbors two well-defined anisotropic superconducting gaps at the Fermi surface, attributed to strongly covalent $\sigma$ bands and $\pi$ bands originating from the $sp$ states of the boron sublattices \cite{an2001, kortus2001, choi2002, mazin2003}.  

Since the discovery of superconductivity in MgB$_2$, researchers have been trying to improve its superconducting properties by alloying. However, MgB$_2$ solid solutions tend to decompose, having a narrow solubility range for most of the dopants, making the formation of high-quality alloys difficult -- except for aluminum. The partial substitution of Mg for Al leads to the loss of superconductivity in MgB$_2$. By preparing Mg$_{1-x}$Al$_x$B$_2$ samples with $0\leq x \leq 0.40$, Ref.~\cite{slusky2001} revealed AlB$_2$-type single-phases for $x$ up to approximately 0.1 and beyond $x = 0.25$. In the intermediate region, two-phase mixtures of AlB$_2$-type were observed as intergrown precipitates on a nanometer scale, suggesting partial collapse of the boron layers separation rather than a change in the B--B in-plane distance \cite{slusky2001}. 

Subsequent studies reported somewhat different results regarding the robustness of bulk superconductivity and the proposed phase transition in Mg-rich phases. Some indicated the complete disappearance of superconductivity only beyond $x=0.7$ \cite{postorino2001,renker2002,li2002}. Ref.~\cite{xiang2002}, for instance, demonstrated that bulk superconductivity persisted up to $x=0.3$ and identified tiny spots of MgAlB$_4$ superstructure in transmission electron microscopy (TEM) images within the low-Al two-phase region (on the order of 0.1\,$\mu\text{m}$). This superstructure was resolved as an ordered phase with doubled lattice parameter along the $c$-axis and alternating Mg and Al layers in the basal plane \cite{xiang2002}. The same was observed in systematic x-ray diffraction studies, TEM images, and energy dispersive x-ray microanalysis \cite{li2002}. Still, the superstructure has been found in a broader composition range (x = 0.1 -- 0.75), likely originating from structural inhomogeneity \cite{li2002}.

The simple picture of an Al--Mg alternately layered superstructure was challenged by high-resolution electron microscopy (HREM), where evidence for a structural modulation within the hexagonal plane was presented, suggesting a more complicated scenario than simple Mg--Al out-of-plane ordering \cite{zandbergen2002}. In the same work \cite{zandbergen2002}, superstructure reflections in electron diffraction only appeared in samples with compositions very close to Mg$_{0.5}$Al$_{0.5}$B$_2$, indicating that the order-disorder transition is likely to be constrained to $x=0.5$ \cite{zandbergen2002}. 

The proposed ordered superstructure around $x = 0.5$, with minor deviations of $\sim 10\,\%$ from the optimal Mg$_{0.5}$Al$_{0.5}$B$_2$ composition, gained further support from synchrotron x-ray powder diffraction \cite{margadonna2002}, and remains the prevailing interpretation to date.

\begin{figure*}[t]
	\includegraphics[width=\linewidth]{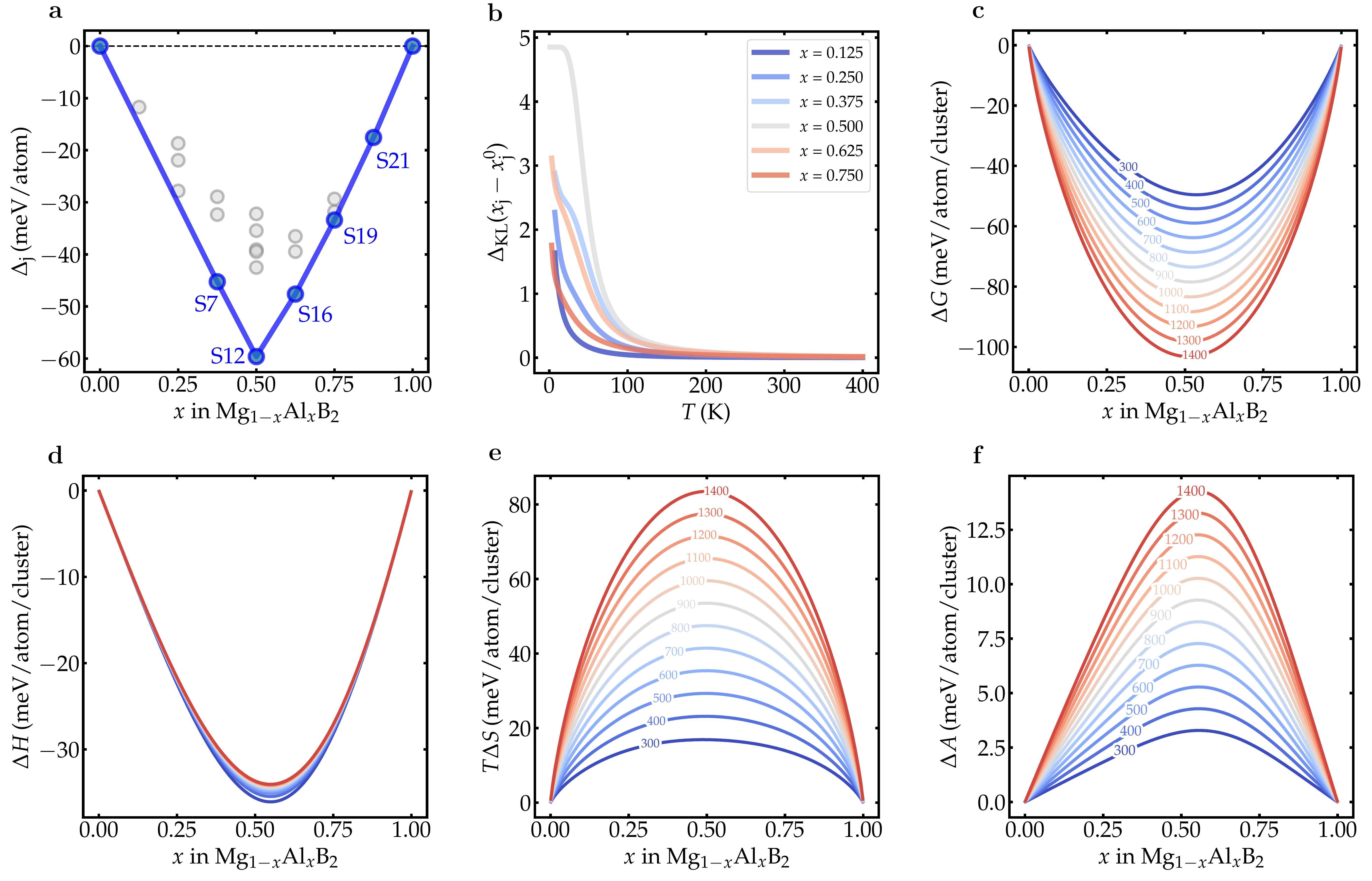}
	\caption{Thermodynamic properties for Mg$_{1-x}$Al$_{x}$B$_2$ solid solution. (a) Excess enthalpy, $\Delta_{j}$, as a function of composition. (b) Kullback-Leibler divergence, $\Delta_{\text{KL}}$, as a function of temperature for selected aluminum compositions below 400\,K. (c) Gibbs mixing free energy, $\Delta G$, (d) Mixing enthalpy, $\Delta H$, (e) configurational entropy, $\Delta S$, and (f) mixing vibrational entropy, $\Delta A$, as a function of composition for crystal growth conditions between 300\,K and 1400\,K.}
	\label{fig:mgalb2_thermo}
\end{figure*}

\textbf{Thermodynamic modeling.} In the case of Mg$_{1-x}$Al$_x$B$_2$, eight non-equivalent atomic sites in the 1$a$ Wyckoff position of the $2\times2\times2$ supercell are occupied either by $N_{\text{A}} = (1-x)N$ magnesium atoms or $N_{\text{B}} = xN$ aluminum atoms. All possible cluster configurations can be reduced to 22 unique classes with distinct total energies $E_{j}$ and degeneracies $g_{j}$, where $j$ = S1, S2, $\ldots$ , S22. 

In (E)GQCA, the supercell size is a parameter to be converged. As demonstrated below, 2$\times$2$\times$2 supercells for Mg$_{1-x}$Al$_x$B$_2$ already show excellent agreement with experimental results. The cluster configurations and their respective degeneracies are detailed in Supplementary Table~2.

We can assess the relative stability of ordered and disordered phases in Mg$_{1-x}$Al$_x$B$_2$ alloys by analyzing the excess enthalpy $\Delta_{j}$, as depicted in Figure~\ref{fig:mgalb2_thermo}(a). In this plot, all 20 off-stoichiometric configurations exhibit negative values, indicating an energy gain relative to the end-members at 0\,K. This suggests a complex competition between a regular solid solution and an order-disorder phase transition led by the clusters lying on the convex hull, \emph{i.e.} S7, S12, S16, S19, and S21, becoming more significant as temperature increases due to the growing entropy contribution.

Among these clusters, S12 represents an ordered phase composed of alternate layers of magnesium and aluminum, with the honeycomb boron lattice sandwiched in between the metallic layers, aligning perfectly with the proposal of Ref.~\cite{xiang2002}. Conversely, S7 and S16 clusters resemble the S12 ordered phase with in-plane point defects at the basal Mg and Al layers, respectively. This picture suggests a pronounced tendency towards ordering in the system, forming Mg--Mg and Al--Al in-plane bonds over a substantial compositional range. Notably, the formation of Al--Mg bonds appears thermodynamically less favorable than the alternate metallic sheets.

The $\Delta_{\text{KL}}(x,T)$ divergence term as a function of temperature for different compositions is depicted in Figure~\ref{fig:mgalb2_thermo}(b) for temperatures below 400\,K. 

The deviations from the random population occur primarily below 300\,K. In the high-temperature regime, the Kullback-Leibler divergence tends towards zero, as expected for proper solid solutions. Nevertheless, for compositions close to $x=0.5$, this divergence is nearly double compared to Mg- and Al-rich samples, even at very high temperatures up to 1200\,K. This discrepancy arises from the cluster energies $\Delta_{j}$ close to the convex hull, suggesting a strong inclination towards clustering. As achieving thermodynamic equilibrium and perfect homogeneous crystal growth conditions is exceedingly challenging in real-world applications, the non-zero $\Delta_{\text{KL}}(x,T)$ divergence at high temperatures suggests a propensity for the precipitation of ordered phases near $x=0.5$ in out-of-equilibrium Mg$_{1-x}$Al$_x$B$_2$ samples.

The deviation of the solid solution from the random cluster distribution can also be assessed from the slightly asymmetric behavior of the mixing enthalpy, $\Delta H(x,T)$, as illustrated in Figure~\ref{fig:mgalb2_thermo}(d). Nonetheless, as the temperature increases, the configurational entropy dominates over the enthalpic contributions, leading to the Gibbs mixing free energy becoming symmetric around $x=0.5$, as seen in Figure~\ref{fig:mgalb2_thermo}(c). 

It is not possible to identify any common tangent line connecting two different $x$ values in $\Delta G$, indicating the absence of spinodal or binodal phase decompositions and, consequently, the eventual formation of a miscibility gap. We want to stress that all these theoretical predictions obtained through EGQCA are consistent with the currently available experimental evidence.

In the case of Mg$_{1-x}$Al$_x$B$_2$, the vibrational free energy, $\Delta A$, plays a minor role in the atomic equilibrium arrangement, even at high temperatures, as shown in Figure~\ref{fig:mgalb2_thermo}(f). At the characteristic growth temperature $T$=1200\,K, for instance, $\Delta A$ contributes approximately 14\,\% to the total Gibbs mixing free energy, with the configurational entropy being much more dominant in the system's thermodynamics. 

It is also important to emphasize at this point that assessing the influence of phonons on the measured physical properties is only possible with the proposed EGQCA. Thus, we believe that our model represents a significant advance in gaining more realistic physical insights, independently of the eventual percentage contribution of $\Delta A$ to $\Delta G$, which can only be evaluated \textit{a posteriori}.

\begin{figure*}[t]
	\includegraphics[width=\linewidth]{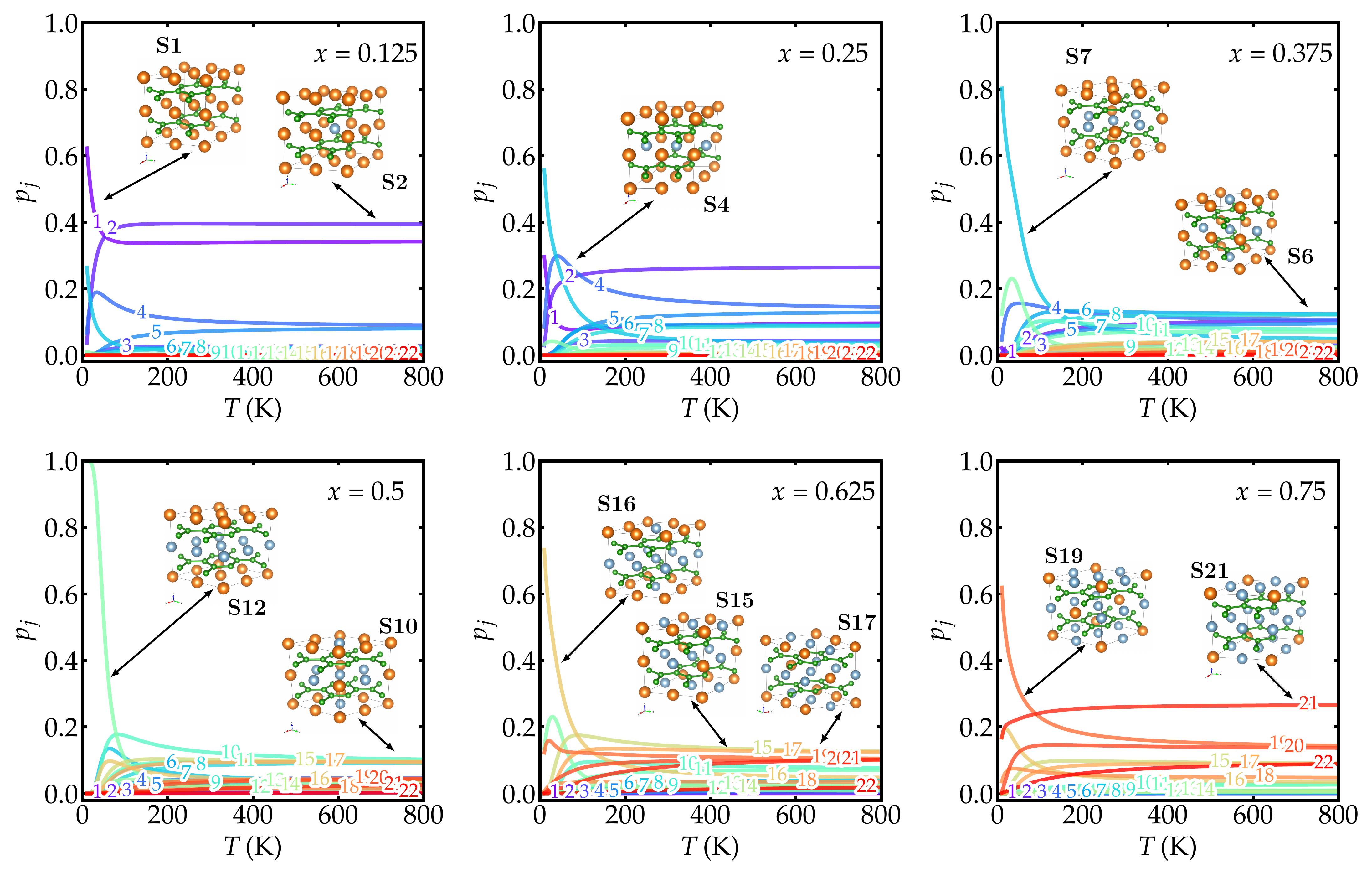}
	\caption{Cluster fraction probabilities for the 22 non-equivalent Mg$_{1-x}$Al$_{x}$B$_2$ clusters as a function of temperature at representative compositions $x = 0.125, 0.25, 0.375, 0.50, 0.625, 0.75$.}
	\label{fig:mgalb2_clusters}
\end{figure*}

\begin{figure*}[t]
	\includegraphics[width=.85\linewidth]{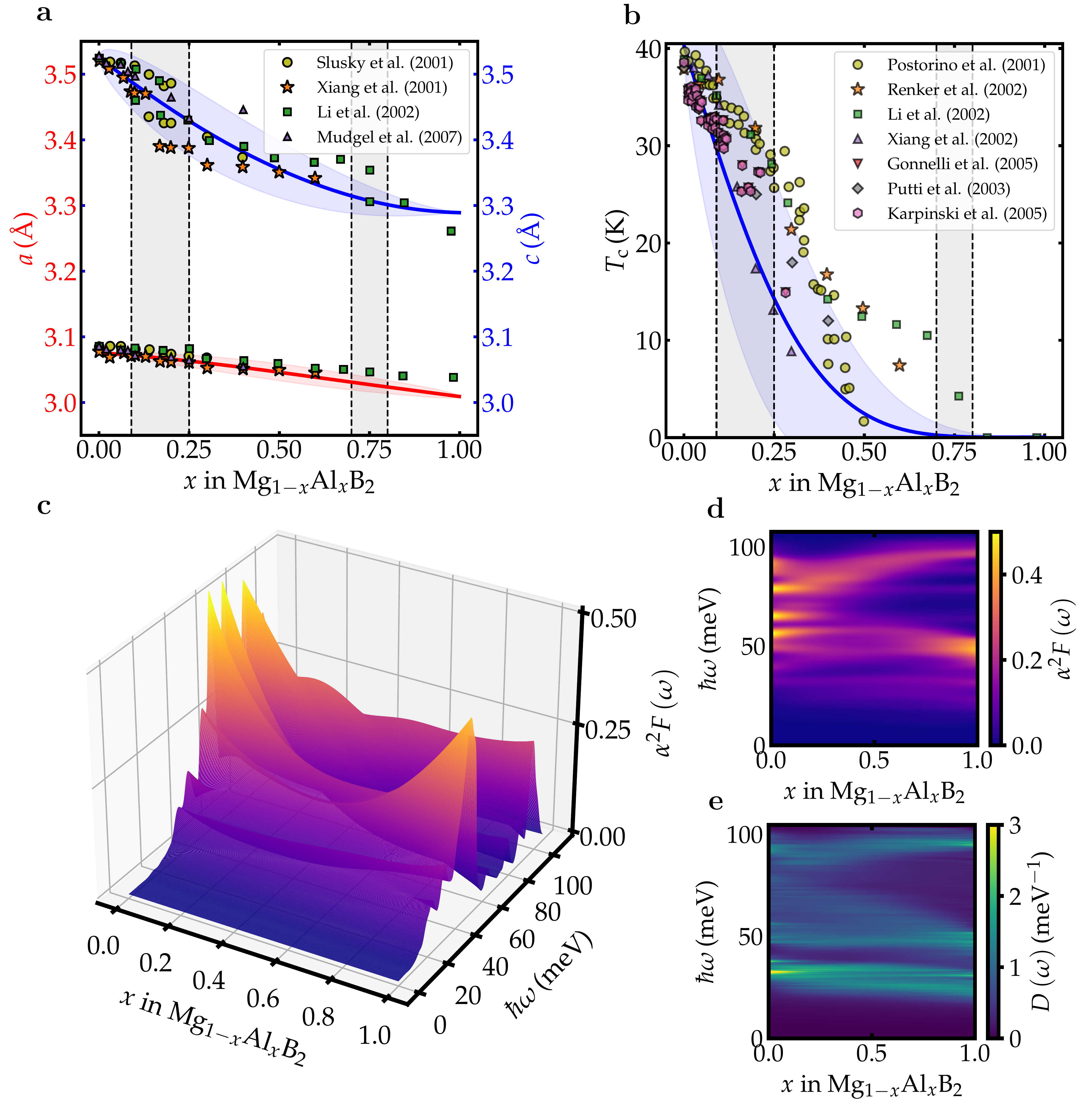}
	\caption{Structural and superconducting properties of Mg$_{1-x}$Al$_x$B$_2$ solid solution within EGQCA: (a) lattice parameters, $a$ and $c$; (b) superconducting critical temperature, {\tc}; (c-d) Eliashberg spectral function, $\alpha^2 F$, as a function of composition and frequency; (e) phonon density of states, $D(\omega)$, as a function of composition and frequency. The solid lines correspond to the configurationally averaged EGQCA values and the shaded areas to the EGQCA standard deviation. Experimental data were extracted from Refs.~\cite{slusky2001, postorino2001, renker2002, xiang2002, li2002, gonnelli2006, putti2003, karpinski2005, mudgel2007}.}
	\label{fig:mgalb2_prop}
\end{figure*}

\textbf{Cluster statistics.} The EGQCA cluster occurrence probability, $p_{j}$, of each cluster class for different compositions is depicted in Figure~\ref{fig:mgalb2_clusters} as a function of temperature. At low temperatures, the ordering tendency is evident. As the aluminum content increases, the system tends to self-organize in order to maximize the probability of finding Al--Al and Mg--Mg bonds, even at stoichiometries far from $x=0.5$. 

At $x=0.25$, for instance, the clusters S2, S4, and S7, which have a probability of finding A--A-type bonds higher than 50\,\%, dominate the atomic arrangement both at low and high temperatures. So, the Gibbs free energy is minimized when in-plane Al--Al and Mg--Mg bonds are formed. 

However, the compromises between the lowest enthalpy, given by the ordered phases, and the randomness induced by entropy tend to create a regular solid solution, even with strong clustering tendencies. 

This competition between enthalpy and entropy is clearly manifested in the microstructures of the Mg$_{1-x}$Al$_x$B$_2$ samples, where tiny spots of MgAlB$_4$ superstructure are preserved in the low-Al two-phase region as intergrown nanometer precipitates \cite{slusky2001, li2002, xiang2002}. The synthesis process also supports our results, as the superstructure intensities depend on the synthesis temperature, being much more visible in the specimens annealed at 800\,$^\circ$C than in those annealed at 900\,$^\circ$C \cite{zandbergen2002}. 

Also, Mg$_{0.5}$Al$_{0.5}$B$_2$ superstructures have demonstrated remarkable stability within the temperature range of 100--500\,K \cite{li2002}, exhibiting no distinguishable changes. Similarly, no thermally induced structural changes were observed in the temperature region between 100 and 700\,K \cite{zandbergen2002}. This suggests that by synthesizing Mg$_{1-x}$Al$_{x}$B$_2$ at very high temperatures, exceeding 1200\,K, and allowing sufficient time for complete relaxation of internal strain energies, thus avoiding the formation of segregates, it might be possible to produce superior single crystals of Mg$_{1-x}$Al$_x$B$_2$ where Al and Mg form regular substitutional solid solutions.    

The cluster fraction probabilities also exhibit remarkable agreement with synchrotron x-ray measurements \cite{margadonna2002}. For instance, our predictions of small fractions of alternately Mg--Al layers with in-plane point defects at low temperatures at $x=0.5$, as indicated by the S7 and S16 clusters in our calculations, align well with observed anisotropic microstrain line-shape broadening to the width of the $\left(001\right)$ and $\left(00\frac{1}{2}\right)$ reflections in Rietveld refinements \cite{margadonna2002}. The predicted small S12 cluster fraction (always smaller than 20\,\%) in the vicinity of $x=0.5$ also corroborates with the diminished intensity and increased width of the $2\theta\approx7.26^\circ$ peak diffraction collected at 16\,K in both Al- and Mg-rich samples, which corresponds to the peak of Mg--Al ordering and superstructure. 

In fact, in synchrotron x-ray diffraction profiles of  Mg$_{0.55}$Al$_{0.45}$B$_2$~\cite{margadonna2002}, the excess Al was found disordered in the Mg layers, consistent with our calculations showing a high S7 cluster fraction in Mg-rich phases near $x=0.5$ (as evident in the representative cluster fraction probabilities in Figure~\ref{fig:mgalb2_clusters} at $x=0.375$). Conversely, for Mg$_{0.45}$Al$_{0.55}$B$_2$, the excess Mg is observed disordered within the Al layers, reflecting the behavior of the S16 cluster at $x=0.625$. 

Therefore, despite the constraints imposed by the supercell size, the EGQCA yields valuable physical insights and sheds light on the outcomes of advanced experimental characterization techniques, such as synchrotron x-ray diffraction. In the Mg$_{1-x}$Al$_x$B$_2$ case, specifically, our thermodynamic modeling aligns with the experimental observations showing that the superstructure formation is likely confined to compositions near to its stoichiometry under conditions of thermodynamical equilibrium. This contrasts with the initial experimental findings~\cite{slusky2001} but is in line with Refs.~\cite{xiang2002,zandbergen2002,margadonna2002}.


\textbf{Physical properties.} The lattice constants $a$ and $c$ of Mg$_{1-x}$Al$_x$B$_2$ as functions of composition $x$ at a fixed growth temperature of 1000\,K are presented in Figure~\ref{fig:mgalb2_prop}(a), obtained using Eqs.~\eqref{eq:EGQCA-props-avg} and \eqref{eq:EGQCA-props-std}. Both lattice parameters decrease with increasing aluminum content in the MgB$_2$ solid solution, which is consistent with the smaller empirical atomic radius of aluminum (1.25\,\AA) compared to magnesium (1.50\,\AA). The lattice mismatch between MgB$_2$ and AlB$_2$ is 2.0\,\% and 1.7\,\% for the hexagonal $a$ and $c$ lattice constants, respectively. The small differences in the equilibrium lattice parameters support the tendency toward a true solid solution under normal growth conditions. Notably, our model exhibits excellent quantitative agreement with various experiments \cite{slusky2001,xiang2002,li2002,mudgel2007}.

The configurationally averaged lattice constant $a$ adheres to Vegard's law~\cite{vegard1921}, which posits that, at constant temperature, there is a linear relation between the alloys' lattice constant and the concentrations of the constituent elements~\cite{danton1991}. The in-plane constant $a$ also presents slight fluctuations dependent on the averaged composition $x$, primarily due to thermal effects, as depicted by the mean-square deviation of the cluster statistics illustrated by the red-shaded area in Figure~\ref{fig:mgalb2_prop}(a). On the other hand, the out-of-plane lattice constant $c$ noticeably deviates from linear behavior concerning the composition $x$, leveling off around $x=0.3$ until reaching the terminal compound composition AlB$_2$. 

One possible physical explanation for the violation of Vegard's law might stem from the increased contribution of the $s$ and $p$ states from Al and Mg at the Fermi level, particularly near the Al-rich side of the solid solution. This enhanced contribution increases the inter-layer coupling of the Mg(Al)-B sheets, consequently shortening the Mg(Al)-B bonds with increasing Al composition. 

While structural fluctuations remain small for the averaged in-plane lattice constant, as revealed by the shaded red area in Figure~\ref{fig:mgalb2_prop}(a), they become non-negligible for the lattice constant $c$. Notably, the standard deviation $\Delta \mathcal{P}(x,T)$ associated with the distribution of the configurationally weighted lattice constant values reflects the accuracy of the experiments: Higher (lower) calculated standard deviations correspond to lower (higher) accuracy in the experimental data. This interpretation of $\Delta \mathcal{P}(x,T)$ underscores, firstly, the robustness of our methodology in describing physical properties of imperfect (real) systems, a notable advantage of EGQCA. Secondly, it underscores the significance of configurational entropy related to the growth conditions in shaping the actual material's physical behavior -- the latter, often overlooked in theoretical predictions and even in the interpretation of measurements. 

The {\tc} for Mg$_{1-x}$Al$_x$B$_2$, obtained using Eq.~(\ref{eq:AD}), is plotted against composition $x$ in Figure~\ref{fig:mgalb2_prop}(b). We adopt a linear interpolation with respect to the composition for the Coulomb pseudopotential between the values corresponding to MgB$_2$ ($\mu^{*}=0.01$) and AlB$_2$ ($\mu^{*}=0.22$). The \mustar{} parameters for the end-members were fitted according to Allen-Dynes equation to reproduce experimental data for \tc{} using calculated $\lambda$, \omegalog{}, and $\overline{\omega}_{2}$. A detailed discussion about \mustar{} within EGQCA and a comparison with computed values is provided later in the text.  

Firstly, it is worth highlighting that achieving a more comprehensive treatment of superconductivity in MgB$_2$ would require a full anisotropic two-band approximation to the electron-phonon coupling and the screened Coulomb repulsion, which differs from our isotropic approach. Nonetheless, even with the most advanced ab initio techniques currently available, further approximations must be made to reproduce the experimentally observed {\tc} value of 39\,K~\cite{margine2013,sanna2020}.

Therefore, we do not aim to quantitatively describe the {\tc} of Mg$_{1-x}$Al$_x$B$_2$ without any adjustable parameter, as this would demand herculean computational efforts. Instead, we strive to accurately describe the general physical behavior, which can be properly assessed through renormalized quantities, while treating the enthalpic, entropic, and vibrational effects on equal footing.

In excellent agreement with experiments, we observe the loss of superconductivity with increasing aluminum concentration. Assuming a regular solid solution, the {\tc} decreases monotonically until approximately $x=0.3$, where a change in slope occurs. The upturn at higher Al content arises from the decrease of $\sigma$ hole-like pockets along the $\Gamma$--$A$ high-symmetry line, leading to an abrupt change in the dimensionality of the Fermi surface dispersion, as discussed in Ref.~\cite{bianconi2002}. The evolution of \nef{}, \omegalog{}, and $\lambda$ as a function of composition is provided in the Supplementary Figure~9.

The agreement between EGQCA and experimental data for {\tc} is noteworthy. Nearly all experimental data points fall within the shaded blue region, which, as previously discussed, is related to the thermal fluctuations and inherent disorder affecting the material's physical properties. The standard deviation associated with the distribution of the weighted {\tc} values indicates the onset of superconductivity loss near $x=0.25$, consistent with the findings in Ref.~\cite{slusky2001}. However, as the system approaches ideal thermodynamic equilibrium, which, experimentally, is directly linked to crystal growth conditions, the complete loss of superconductivity is expected just at $x=0.75$, as reported in Ref.~\cite{li2002}. This suggests that discrepancies among various experiments likely stem from differing thermodynamic conditions during crystal growth.

We want to note that our model represents the first one to quantitatively capture the {\tc} behavior of Mg$_{1-x}$Al$_x$B$_2$ without the need of performing electron-phonon Eliashberg function scaling \cite{kortus2005} or solely taking the zone-center mode \cite{de2002,profeta2003,sun2022}. Although Ref.~\cite{de2010} achieved a good agreement up to $x=0.2$ by solving the two-band Eliashberg gap equations, they employed the VCA model. The main concern with studying doping effects in superconducting systems using VCA lies in the uncertainty surrounding mixing different electronic configurations, thereby questioning the general validity of the electron-phonon coupling strength description. The great qualitative success of VCA in describing the superconducting properties of electron- and hole-doped MgB$_2$ alloys \cite{de2002,profeta2003,de2010} hinges on the significance of the in-plane stretching boron modes and $\sigma$ bands. Nevertheless, this is generally not applicable to other superconductors. Hence, a more rigorous and robust treatment requires the summation over all phonon modes across the entire Brillouin zone, a feature intrinsically addressed in EGQCA.

It is noteworthy that even without considering anisotropic or anharmonic effects, we can capture the primary features of the Mg$_{1-x}$Al$_x$B$_2$ solid solution using EGQCA. These features include the order-disorder tendency, the atomic arrangements revealed by characterization techniques, and the evolution of {\tc} with varying Al content. This approach helps reconcile discrepancies observed in different experiments. In future studies, it would be interesting to compare the EGQCA using the anisotropic Migdal-Eliashberg approach~\cite{margine2013,lucrezi2024} with the results presented in this paper.

\textbf{Electron-electron interactions.} To demonstrate that linearly interpolating the \mustar{} obtained from the existing experimental measurements or calculating them fully ab initio provides similar outcomes, we treated the fully screened Coulomb interactions using the \textit{GW} approximation \cite{giustino2010,lambert2013} for MgB$_2$ and AlB$_2$. The electron-electron interaction strength is obtained as $\mu$ = \nef{}$ \left<\left<V_{\text{k},\text{k}^{'}}\right>\right>_{\text{FS}}$, where $V_{\text{k},\text{k}^{'}} = \left<\mathbf{k}, -\mathbf{k} \left| W \right| \mathbf{k^{'}}, -\mathbf{k^{'}} \right>$, and $W$ is the screened Coulomb interaction within the random phase approximation~\cite{lee1995}. The Morel-Anderson pseudopotential~\cite{morel1962} is then computed as \mustar{} = $\mu/\left[1 + \mu\ln\left(\omega_{\text{el}}/\omega_{\text{ph}}\right)\right]$, with $\omega_{\text{el}}$ and $\omega_{\text{ph}}$ being the characteristic electron and phonon energies, respectively. For use within the Allen-Dynes formalism, $\omega_{\text{ph}}$ is in the range of the Debye frequency, while within Eliashberg formaism, the corresponding phonon scale is in the range of the Matsubara cutoff frequency~\cite{flores2020}.

Our calculations yield a Coulomb pseudopotential of $\mu^{*}=0.09$ for MgB$_2$ and $\mu^{*}=0.14$ for AlB$_2$ to be used within Allen-Dynes formalism. Hence, for off-stoichiometric compounds, we interpolated $\mu^{*}$ linearly based on the \textit{GW} calculated values for the terminal constituents. The results are shown in Supplementary Figure~10. Notably, there is almost no difference between the normalized \tc{} of Mg$_{1-x}$Al$_x$B$_2$ as a function of the composition obtained through approximations $(i)$ and $(ii)$, except for a slight upturn in \tc{} close to AlB$_2$ for the \mustar{} derived from the \textit{GW} approximation. This deviation arises because a substantially higher Coulomb pseudopotential is necessary to fully suppress the superconductivity of aluminum diboride. Hence, leveraging available experimental information to ascertain a more consistent \mustar{} behavior yields comparable results to the fully ab initio approach, with the great advantage of saving considerable computational time.

\subsection{Nb-alloys: addressing open questions}
\label{sec:nb}

\textbf{Superconducting niobium alloys}. Niobium has been the fundamental building block of superconducting devices such as superconducting radio-frequency (SRF) cavities for particle accelerators~\cite{gurevich2012} and SQUIDs for sensing magnetic fields on microscopic and nanometric scales~\cite{schmelz2015}. Nb-Ti alloys have been particularly popular since the early 1980s, being used in the first superconducting particle accelerator and the first commercial magnetic resonance imaging (MRI) machine~\cite{scalan2004}. Due to their high upper critical field and good ductility, among other reasons, they have been the workhorse of the superconductor industry and have maintained this status to date.

We considered Nb--Ti and Nb--V alloys as promising test beds for the EGQCA due to their application relevance and abundant experimental data~\cite{hulm1961}. From a theoretical point of view, Nb--Ti, Nb--V, or even pure niobium are incredibly intricate systems, as we will demonstrate in the following. 

For describing the Nb-alloys within (E)GQCA, we adopted a 2$\times$2$\times$1 supercell of the niobium's body-centered cubic ($bcc$) 2-atom unit cell, which gives 256 possible configurations in the full disordered case. By identifying the degenerate configurations, we end up with 22 non-equivalent clusters, for which we fully relaxed the cell parameters and atomic degrees of freedom while keeping the Bravais-lattice index of the supercell fixed. The cluster's information and physical properties are listed in Supplementary Tables~3-4.

\textbf{Nb--Ti solid solution}. The Nb--Ti case is very instructive for discussing how to handle situations where the alloyed system contains dynamically unstable clusters or, more drastically, as in the Nb$_{1-x}$Ti$_x$ solid solution, where one of the end members lacks dynamic stability at 0\,K. 

Titanium is hexagonal ($\alpha$ phase) under ambient conditions but becomes $bcc$ ($\beta$ phase) at high temperatures. The phase diagram of Nb--Ti is extremely intricate~\cite{zhang2001}, especially in the vicinity of the Ti-rich region, and the superconductivity of the Nb$_{1-x}$Ti$_x$ solid solution is markedly influenced by the sample's heat treatment~\cite{hulm1961}. According to Ref.~\cite{hulm1961}, the solubility limit of the cubic phase during cooling the samples towards 0\,K is approximately Ti-80\,at.\%, where the high-temperature $\beta$ phase undergoes a diffusionless transformation to the $\alpha$ phase~\cite{hulm1961}. 

\textbf{Dealing with dynamically unstable clusters}. Indeed, the $\beta$-Ti at 0\,K is highly unstable within the harmonic approximation according to our calculations, with several imaginary phonon modes across the whole Brillouin zone, as illustrated in the Suplementary Figure~13. We computed the phonon dispersion and phonon DOS for all 22 non-equivalent clusters of the Nb$_{1-x}$Ti$_x$. All clusters with composition exceeding Ti-62.5\,at.\% are dynamically unstable, encompassing up to 8 out of the 22 clusters. More details can be found in the Supplementary Table~3. 

In particular, the $\beta$-Ti phase, one of the end members of the Nb$_{1-x}$Ti$_x$ solid solution, is highly unstable at 0\,K within the harmonic approximation. In such instances, incorporating the vibrational mixing free energy into the GQCA formalism is not possible, as the reference $\Delta A$ would be substantially underestimated due to entire modes being absent from the phonon DOS integration. That is why we employed the usual GQCA for Nb--Ti, omitting vibrational effects on the free energy.

One can still question, though, if excluding the imaginary phonon modes for computing the \af{}, as discussed in Section~\ref{sec:gqca}, will not underestimate the \tc{}. With this in mind, we integrated the phonon DOS from zero up to the Debye frequency for all clusters to compute the percentage of excluded modes. The results are detailed in the Supplementary Tables~2-5. For Nb--Ti, only 4 out of the 8 dynamically unstable clusters have more than 5\,\% of the modes excluded by integrating over $\omega \geqslant 0$. Solely one configuration, the $\beta$-Ti, has over 10\,\% of modes discarded. Since (E)GQCA averages over the cluster fraction probabilities at every temperature and composition, we argue that the excluded phonon modes have a negligible impact on the description of physical properties. 

To illustrate, we examine the cluster fraction probabilities at $x=0.75$, a region close to the solubility limit of the $\beta$ phase in Nb--Ti, where unstable clusters should notably contribute to the solid solution. At $T=800$\,K, a temperature regime where the Kullback-Leiber divergence approaches zero (see Supplementary Figure~13), the imaginary phonon modes represent only 5\,\% of the total number of modes contributing to the alloy's phonon-derived properties at that composition. Thus, the influence of these modes on the final results is effectively mitigated by GQCA's cluster averaging scheme, justifying the approximation carried out for Nb--Ti.   

\begin{figure}[t]
	\includegraphics[width=\linewidth]{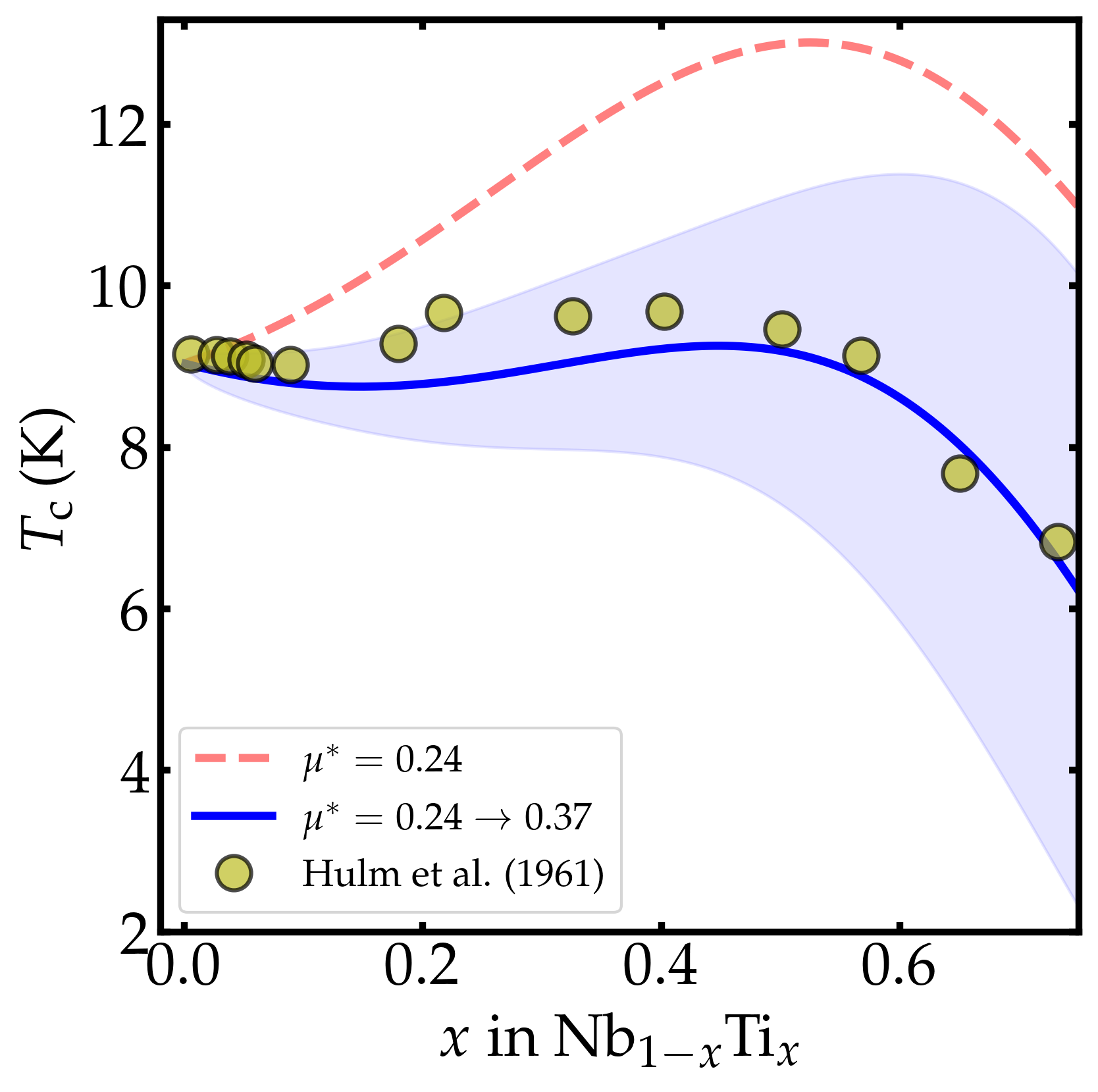}
	\caption{Superconducting critical temperature, {\tc}, of Nb$_{1-x}$Ti$_x$ solid solution as a function of composition within GQCA. The  lines correspond to the configurationally averaged GQCA values and the shaded area to the GQCA standard deviation. The dashed red line represents \tc{} obtained with \mustar{} = 0.24 for all clusters, whereas the solid blue line employs linear interpolation from 0.24 ($x = 0.00$) to 0.37 ($x = 0.75$). Experimental data were extracted from Ref.~\cite{hulm1961}.}
	\label{fig:nbti_tc}
\end{figure}

\textbf{Nb--Ti Superconducting properties}. The \tc{} of Nb$_{1-x}$Ti$_x$ as a function of composition within GQCA is shown in Figure~\ref{fig:nbti_tc}. In the figure, the dashed red line represents GQCA using \mustar{} = 0.24 for all clusters, whereas the solid blue line employs linear interpolation from 0.24 to 0.37. The \mustar{} parameters for pure Nb and Nb-Ti alloy at the solubility limit were fitted using Allen-Dynes formula to replicate the experimental data for \tc{}, utilizing the calculated $\lambda$, \omegalog{} and $\overline{\omega}_{2}$. 

Electron-electron interactions play a pivotal role in reproducing the experimental data reported in Ref.~\cite{hulm1961}, denoted by yellow solid circles in Figure~\ref{fig:nbti_tc}. By assuming an increasing Coulomb pseudopotential for Nb--Ti relative to the titanium solute, we achieve a remarkable agreement with experiments. Even nuanced features, such as the slight upturn of \tc{} in the Nb-rich region and the rapid decrease beyond Ti-50\,at.\% up to the martensite transformation, are captured by the GQCA formalism.      

Regarding the Morel-Anderson pseudopotential, the calculated values for \mustar{} in niobium from several different methods (full ab initio calculations and experiments) vary from 0.18 to 0.26~\cite{scott1970,butler1979,zarea2023}. Our estimated value of 0.24 for \mustar{}, fitted according to Allen-Dynes equation to reproduce the experimental \tc{} using the calculated $\lambda$, \omegalog{}, and $\overline{\omega}_{2}$, fully agrees with the value extracted from Hass–van Alphen data used to reconstruct the Fermi surface in niobium~\cite{scott1970}. The calculated values for $\lambda$ in niobium, on the other hand, vary from 0.58 to 1.33~\cite{butler1977,butler1979,crabtree1979,bauer1998,davydov2020,kawamura2020,zarea2023}, with most of them around $\lambda = 1.2$, which is consistent with the value of $\lambda = 1.33$ found in this work.

The results in Figure~\ref{fig:nbti_tc} should be interpreted cautiously though. The $bcc$ Nb--Ti solid solution is not homogeneous in practical applications. Depending on the cooling method employed during alloy synthesis, one may obtain a regular solid solution, a highly inhomogeneous Nb-rich $\beta$ phase, or martensite precipitation of the $\alpha$ phase with the same composition as the parent $\beta$ phase~\cite{hulm1961}, and the degree of martensite precipitation or impurities directly impacts the superconducting properties~\cite{witcomb1973,chernyj1991}. However, (E)GQCA offers a robust and elegant way to assess the proximity of experimental data to an ideal, regular solid solution, \emph{i.e.} free of solid reactions, nucleations, flux pinning effects, oxides and sub-oxides precipitates, and impurities.

For Nb--Ti, it is possible to accurately describe the available \tc{} data by assuming that electron-electron renormalization effects increase linearly with rising titanium content. The GQCA can even capture the more complex features of the evolution of \tc{} with composition, which is a remarkable achievement. At the present moment, it is not fully understood if these features are indeed caused by the enhancement of the Coulomb pseudopotential in the Ti-rich portion or by microstructure effects, but we expect our work will stimulate further experimental and theoretical endeavors to fill this gap. 

During the preparation of this manuscript, ab initio calculation of the superconducting properties for the ordered, equiatomic $\beta$-NbTi within the full-bandwidth anisotropic Migdal-Eliashberg formalism~\cite{lucrezi2024} incorporating quantum anharmonic effects has been reported in Ref.~\cite{cucciari2024}. They determined a \mustar{} of 0.20 according to the Morel-Anderson approximation and obtained a \tc{} of 23.7\,K~\cite{cucciari2024} at $x = 0.5$. Employing rescaling of \nef{} based on Boltzmann-averaged supercell model to account for disorder, the \tc{} decreased to 18.9\,K, which is still considerably higher than the experimental values. These results support our view that Nb--Ti superconductivity is a highly intricate case of conventional superconductor. Even when including quantum anharmonic lattice effects, energy-dependent Coulomb interactions, spin fluctuations, and using the Eliashberg formalism beyond the Fermi surface, as demonstrated in Ref.~\cite{cucciari2024}, the predicted \tc{} significantly exceeds experimental values.

It is important to remember that the \mustar{} used within Eliashberg theory should be rescaled compared to the one when solving the Allen-Dynes equation, as Matsubara cutoff is the relevant energy scale~\cite{allen1975}. Specifically, by rescaling the Eliashberg \mustar{} of 0.20 used in Ref.~\cite{cucciari2024} for $\beta$-NbTi to 0.145 according to Ref.~\cite{allen1975}, GQCA predicts a \tc{} of 18.2 K. However, our calculations do not account for anharmonic effects, which are essential for accurately describing the phonon dispersions and stability of the ordered $\beta$-NbTi~\cite{cucciari2024}. If it is computationally feasible, including these effects is expected to significantly decrease the GQCA-predicted \tc{}, aligning GQCA more closely with experimental results without needing a substantial renormalization of the Coulomb pseudopotential.

\textbf{Spin-fluctuation suppressed superconductivity in Nb--V}. The Nb--V case poses even greater challenges from a theoretical perspective compared to Nb--Ti. According to Ref.~\cite{kawamura2020}, vanadium exhibits the largest Coulomb repulsion and spin-fluctuation kernel ($\mu_{\text{SF}}$ = 0.722) among all elemental superconductors, followed by niobium ($\mu_{\text{SF}}$ = 0.203) and tantalum ($\mu_{\text{SF}}$ = 0.131).   

Ref.~\cite{tsutsumi2020} performed SCDFT calculations incorporating plasmon and spin fluctuations for vanadium and niobium, quantifying the effects of paramagnons on the \tc{} without any empirical parameters.  Their findings reveal that plasmons enhance vanadium's \tc{} from 7.3\,K to 13.1\,K. However, further inclusion of spin fluctuations suppresses \tc{} to 1.8\,K. The relatively strong ferromagnetic-fluctuation-depressed \tc{} in vanadium can be tentatively explained by the localized $3d$ valence states of vanadium, where Pauli exchange effect holds greater significance.

Ref.~\cite{tsutsumi2020} also estimated \mustar{} for the Allen-Dynes formalism, similar to our approach. They calculated $\lambda$, \omegalog{}, and $\overline{\omega}_{2}$ applying DFPT, and the \tc{} using SCDFT, and then utilized these values to reproduce the same SCDFT-\tc{} with the Allen-Dynes formula. The fitted \mustar{} values, based on the calculated SCDFT-\tc{} including spin fluctuations, are 0.419 for V and 0.277 for Nb. Consequently, Ref.~\cite{tsutsumi2020} concludes that conventional \mustar{} values for Allen-Dynes equation ranging from 0.1 to 0.2 do not apply to vanadium and niobium due to the unusually strong ferromagnetic fluctuations in these materials. 

To reproduce the experimental \tc{} values for V and Nb using the Allen-Dynes formula, we thus adopted similar fitted \mustar{} values of 0.30 and 0.24, respectively. Although considerably higher than typical values for the Coulomb pseudopotential predominantly observed in BCS superconductors, these values, as discussed earlier, align perfectly with the available experimental and theoretical data, and effectively decrease the electron-phonon mass enhanced parameter due to the effects of spin fluctuations, which are not accounted for in our calculations.

The spin-fluctuation effects are also known to play a pivotal role in Nb--V alloys. Ref.~\cite{rapp1974} estimated the contributions of spin fluctuations to the suppression of \tc{} in Nb$_{1-x}$V$_x$ solid solutions from the electrical resistivity conducted around 250\,ºC along with published data on susceptibility and the electronic part of the specific heat. To reconcile the measured \tc{} values with those predicted by the McMillan formula~\cite{mcmillan1968}, they employed a linear interpolation of \mustar{} from 0.13 (pure niobium) to 0.18 (pure vanadium). The significance of spin fluctuations in determining \tc{} is further supported by the mass enhancement factor obtained from band structure calculations, the observed anomalous spin-lattice relaxation time, and the enhanced spin susceptibility~\cite{rapp1974}.

\begin{figure*}[t]
	\includegraphics[width=\linewidth]{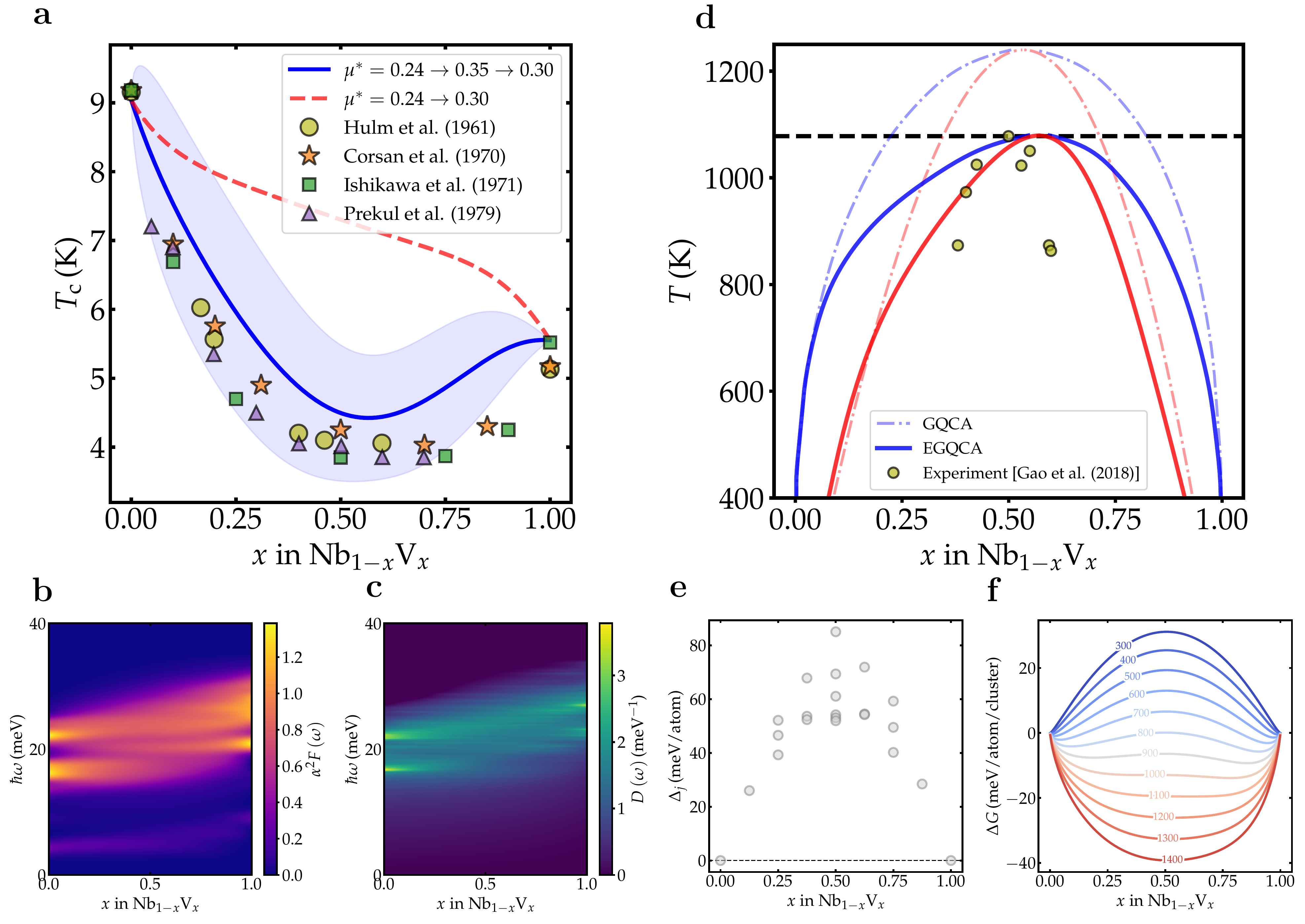}
	\caption{Thermodynamics and superconducting properties of Nb$_{1-x}$V$_x$ solid solution within EGQCA. (a) superconducting critical temperature, {\tc}, according to the Allen-Dynes equation. The dashed red line employs a linear interpolation of \mustar{} from 0.24 ($x = 0.00$) to 0.30 ($x = 1.00$), and the solid blue line employs a linearly interpolated \mustar{} from 0.24 ($x = 0.00$) to 0.35 ($x = 0.50$) to 0.30 ($x = 1.00$); (b) Eliashberg spectral function, $\alpha^2 F$, as a function of composition and frequency; and (c) phonon density of states, $D(\omega)$, as a function of composition and frequency. (d) Miscibility gap: binodal (blue) and spinodal (red) phase decompositions according to EGQCA (solid lines) and GQCA (dashed lines); (e) excess energy, $\Delta_{j}$; (f) Gibbs mixing free energy, $\Delta G$, including vibrational contributions between 300\,K and 1400\,K;  The lines in (a) correspond to the configurationally averaged EGQCA values and the shaded area to the EGQCA standard deviation. The experimental data were extracted from Refs.~\cite{hulm1961,corsan1970,ishikawa1971,prekul1979,gao2018}.}
	\label{fig:nbv_prop}
\end{figure*}

\textbf{EGQCA}. Interestingly, assuming a linear behavior of \mustar{} as a function of composition, as employed for Mg$_{1-x}$Al$_x$B$_2$, Nb$_{1-x}$Ti$_x$, and in Ref.~\cite{rapp1974}, fails to replicate the experimental trend for \tc{} for Nb$_{1-x}$V$_x$ within the EGQCA framework. The electron-phonon mass enhancement parameter is unable to account for the minimum observed in the \tc{} curve around $x$ = 0.5, as the electron-phonon coupling decreases with increasing vanadium content. To reconcile the experimental values reported in Refs.~\cite{hulm1961, corsan1970,ishikawa1971,prekul1979}, an additional adjustable \mustar{} must be included, as depicted in Figure~\ref{fig:nbv_prop}(a). A renormalized \mustar{} of 0.35 at $x = 0.5$ is necessary to suppress superconductivity to around 4.5\,K. Hence, our calculations suggest that disordered, off-stoichiometric Nb$_{1-x}$V$_x$ compounds could exhibit even stronger spin fluctuations than pure vanadium or niobium. 

However, the nature of such spin fluctuation enhancement is not entirely clear, particularly as \nef{} rises steadily with increasing vanadium fraction. An alternative hypothesis is that, similar to ordered NbTi, quantum anharmonic lattice effects should be taken into account in disordered Nb$_{1-x}$V$_x$ configurations to renormalize the Eliashberg spectral function. This represents an interesting avenue for future investigations. Still, it is noteworthy that by employing only three adjustable parameters --- \mustar{} at $x$ = 0.0, 0.5, and 1.0 --- we are capable to capture the intricate behavior of \tc{} as a function of composition for Nb$_{1-x}$V$_x$ across the entire solubility range.     

\textbf{Miscibility gap}. The Nb$_{1-x}$V$_x$ solid solution, on the other hand, is a representative system to demonstrate the advantages of the EGQCA formulation compared to GQCA.   

Figure~\ref{fig:nbv_prop}(e) and (f) depict the excess enthalpy, $\Delta_{j}$, and the resulting Gibbs mixing free energy, $\Delta G$, obtained for Nb$_{1-x}$V$_x$. All $\Delta_{j}$ values for the 20 off-stoichiometry clusters considered have positive energy, indicating the high stability of the end members and the need for high temperatures to stabilize the solid solution phase. The $\Delta G$ curves for different temperatures exhibit marked asymmetry concerning composition. Below 1078\,K, $\Delta G$ displays a common tangent at two distinct $x$ values, indicating the alloys' tendency to phase decomposition. 

For temperatures lower than 1078\,K, referred to as the critical temperature of the miscibility gap, $T_\text{mg}$, there exists a region in parameter space $x-T$ (miscibility gap) where the alloy becomes unstable and tends to segregate into two distinct stable phases with concentrations $x_1$ and $x_2$, such that, at thermodynamic equilibrium, both phases share the same chemical potentials. The equilibrium $x_1$ and $x_2$ compositions, at a given temperature, that define this so-called binodal equilibrium compositions, can be determined by standard methods in thermodynamic optimization, corresponding to one or more common tangent constructions, resulting in the convex hull at the given temperature~\cite{Lee1992a}.
In the present work, the Quickhull algorithm, as implemented in the Qhull library  \cite{Barber1996}, was employed for this purpose, resulting in the binodal curves shown in Figure~\ref{fig:nbv_prop}(d).

The spinodal lines in Figure~\ref{fig:nbv_prop}(d) correspond to any inflection points in the free energy curve at a given temperature, i.e., the compositions $x_s$ for which 
\begin{align}
	\left . \dfrac{\partial^2\Delta G(x,T)}{\partial^2 x} \right|_{x=x_s} = 0\,.
\end{align}
If the concavity of $\Delta G(x,T)$ is negative, small perturbations tend to lower the free energy, and the system decomposes into two distinct phases (with compositions given by the binodal lines) in order to reach the lowest free energy state possible. Conversely, even if positive, and the fact that decomposition into phases with composition $x_1$ and $x_2$ could lower the global free energy, small perturbations will locally increase the Gibbs mixing free energy, creating an energy barrier for decomposition. Therefore, a single-phase solution with composition inside the spinodal region in Figure~\ref{fig:nbv_prop}(d) is intrinsically unstable and will decompose in two phases, whereas a solution with composition between the binodal and spinodal regions will be stable under small perturbations. 

We calculated the spinodal (red lines) and binodal curves (blue lines), as illustrated in Figure~\ref{fig:nbv_prop}(d), utilizing both GQCA (dashed lines) and EGQCA (solid lines). Experimental data for spinodal decomposition (when the alloy composition lies within the spinodal dome) and the nucleation-growth decomposition (when the alloy composition falls between the binodal and spinodal curves)~\cite{gao2018} are represented as yellow points in Figure~\ref{fig:nbv_prop}(d). GQCA significantly overestimates the critical temperature of the miscibility gap, yielding a $T_{\text{mg}}$ of $\approx$ 1250\,K. Conversely, EGQCA excellently matches the experimental value of 1078\,K determined in Ref.~\cite{gao2018} through differential thermal analysis (DTA).  Furthermore, the predicted solubility ranges are much narrower within the EGQCA framework. For a crystal growth temperature of 1000\,K, for instance, the solubility limit of V in Nb according to GQCA is less than 20\,\%, while EGQCA correctly predicts a solubility limit exceeding 30\,\%. Such a picture is in agreement with the experimental findings~\cite{gao2018}. Thus, our results indicate that vibrational entropy is the driving force for forming the actual miscibility gap of the $bcc$ phase in the Nb--V system, acting to decrease the critical temperature and increase the solubility limit of the $bcc$ structure. 

The description of the miscibility gap for Nb--V by EGQCA notably surpasses previous theoretical endeavors~\cite{ravi2012,kumar2022,wang2023}. The most recent and comprehensive attempt to theoretically describe the miscibility gap of the $bcc$ phase in Nb--V is presented in Ref.~\cite{wang2023}, where chemical order, lattice distortion, and lattice vibrations were incorporated into the calculations by combining cluster expansion and Monte Carlo simulations. Although they predict the correct critical temperature for the miscibility gap, the regions of phase decomposition are overestimated, predicting a much lower solubility of Nb in V and vice versa below the critical temperature. Additionally, the methodology employed doesn't enable the calculation of the spinodal curve. 

EGQCA stands out as the sole method capable of describing the marked asymmetry of the miscibility gap, which is even more pronounced in DTA measurements. We conclude that EGQCA can accurately predict the critical temperature, offer a more precise description of the solubility limit, and capture the asymmetry of the miscibility gap using only 22 clusters, showcasing the power and effectiveness of EGQCA in accurately describing the thermodynamics of solid solutions.

\subsection{Designing high-{\texorpdfstring{\tc}{Tc}} alloys: \texorpdfstring{Y$_{1-x}$Ca$_x$H$_6$}{Y(1-x)Ca(x)H6}}
\label{sec:ycah6}

\textbf{Alloyed superhydrides}. The discovery of the compressed superconducting hydrogen sulfide at 203\,K under 145\,GPa in 2015~\cite{drozdov2015} has revived the search for high-temperature superconductivity (HTSC). Since then, numerous superhydrides have been synthesized~\cite{flores2020,pickard2020,gao2021,lilia2022}. HTSC has been successfully identified in the sodalite-like clathrates LaH$_{10}$~\cite{drozdov2019,semenok2022}, CeH$_{10}$~\cite{chen2021}, CeH$_{9}$~\cite{chen2021}, YH$_9$~\cite{kong2021}, YH$_6$~\cite{kong2021,troyan2021}, and CaH$_6$~\cite{li2022,ma2022}, and predicted in many other systems~\cite{zhao2023}.

Doping stands out as one of the most promising routes to lower the stabilization pressure of binary and ternary superhydrides while maintaning the strong electron-phonon coupling of the hydrogen rich phases. Recently, a series of (La,Y)H$_{10}$ ternary hydrides were experimentally obtained at pressures of 170--196\,GPa exhibiting a maximum \tc{} of 253\,K. Solid solutions of $P6_3/mmc$-(La,Ce)H$_{9-10}$ were also synthesized~\cite{bi2022,chen2023}. The superconductivity of the $P6_3/mmc$ phase was observed down to approximately 100\,GPa with \tc{}~=~176\,K, marking a significant enhancement of the superconducting properties compared with the binaries La--H and Ce--H in the same pressure range~\cite{chen2023}. 

The experimental observation of \tc{} enhancement in $P6_3/mmc$-(La,Ce)H$_{9-10}$ solid solutions has demonstrated that the formation of alloys contributes not only to reducing the stabilization pressure but also to designing superior superconducting properties~\cite{bi2022,chen2023}. The highest \tc{} predicted in ternary hydrides, for instance, is attributed to the clathrate structure Li$_2$MgH$_{16}$, presenting a remarkably \tc{} of 473\,K at 250\,GPa~\cite{sun2019}. Li$_2$MgH$_{16}$ was obtained by simply doping the low-\tc{} binary superconductor MgH$_{16}$ with Li atoms to dissociate H$_2$ molecules into a high amount of pure hydrogen-free states at the Fermi level. 
\begin{figure*}[t]
	\includegraphics[width=\linewidth]{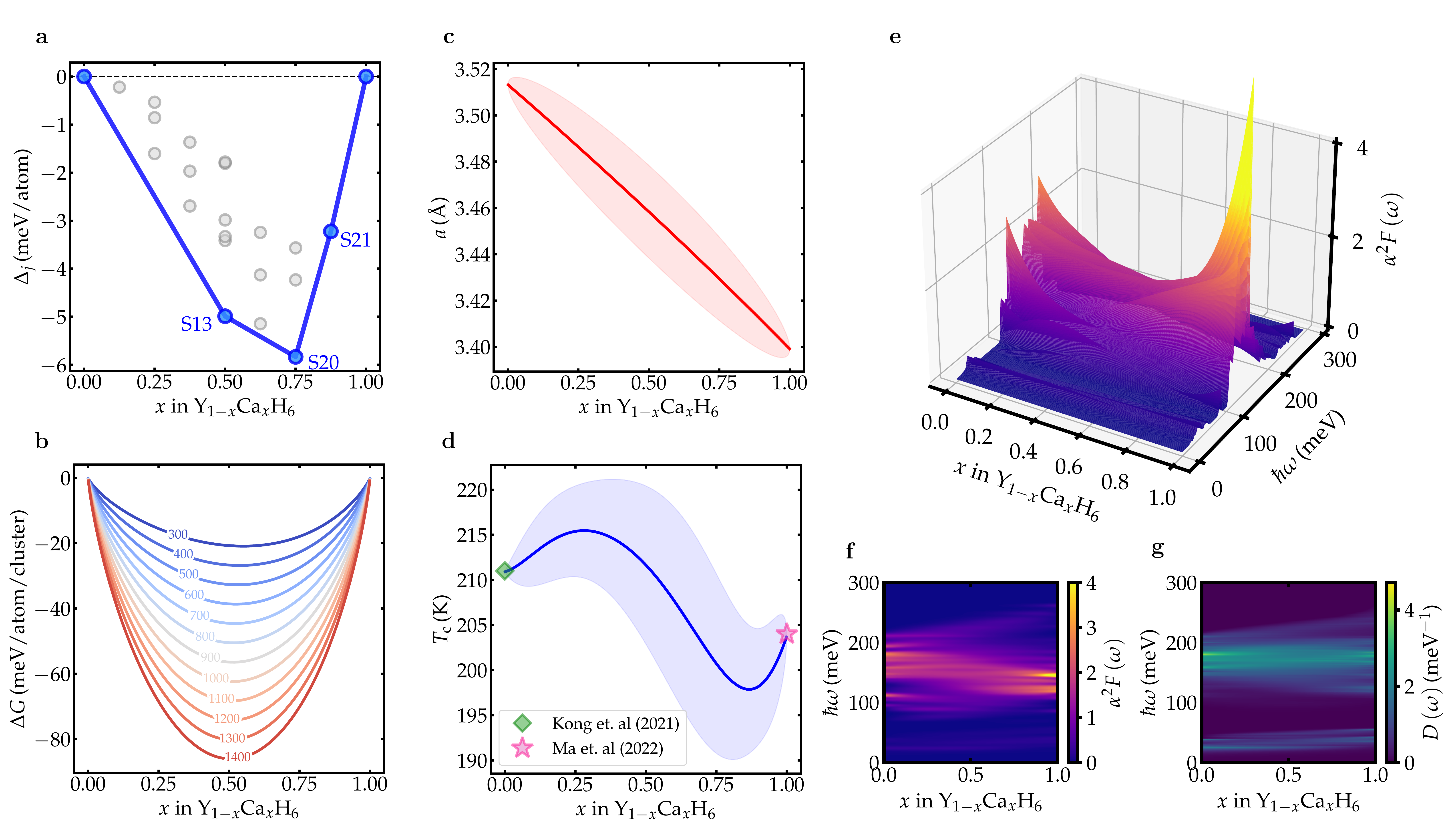}
	\caption{Thermodynamics and physical properties of Y$_{1-x}$Ca$_x$H$_6$ solid solution at 200\,GPa within EGQCA: (a) excess enthalpy, $\Delta_{j}$; (b) Gibbs mixing free energy, $\Delta G$, including vibrational contributions between 300\,K and 1400\,K; (c) lattice parameter, $a$; (d) superconducting critical temperature, {\tc}, according to the machine-learned UF formula; (e-f) Eliashberg spectral function, $\alpha^2 F$, as a function of composition and frequency; (g) phonon density of states, $D(\omega)$, as a function of composition and frequency. The solid lines in (b), (c), and (d) correspond to the configurationally averaged EGQCA values and the shaded areas in (c) and (d) to the EGQCA standard deviation. Experimental data were extracted from Refs.~\cite{kong2021,ma2022}.}
	\label{fig:ycah6_prop}
\end{figure*}

In this vein, EGQCA is a powerful tool for designing new alloyed superhydrides. To demonstrate this, we have, for the first time, modeled the Y$_{1-x}$Ca$_x$H$_6$ solid solution. The YH$_6$--CaH$_6$ system adheres to almost all the Hume-Rothery rules~\cite{hume1934} for forming a regular substitutional solid solution experimentally: $(i)$ the percentage difference of the atomic radii between Ca and Y is less than 15\,\%; $(ii)$ both end members possess the same crystal structure, are dynamically stable, and have been confirmed experimentally; $(iii)$ they have similar electronegativity. As the high-pressure synthesis of hydrides has evolved significantly in the past few years~\cite{eremets2024}, synthesizing high-quality Y$_{1-x}$Ca$_x$H$_6$ solid solutions should therefore be feasible.  

\textbf{Thermodynamics}. For modeling the Y$_{1-x}$Ca$_x$H$_6$ at 200\,GPa within EGQCA, we used a  2$\times$2$\times$1 supercell, resulting in a total of 256 configurations and only 22 non-equivalent clusters, for which we computed the structural, electronic, and superconducting properties, as summarized in Figure~\ref{fig:ycah6_prop}. All off-stoichiometric clusters are dynamically stable and exhibit an energy gain relative to the end-members at 0\,K, as evidenced by their negative excess enthalpies, with 3 clusters lying on the convex hull, \emph{i.e.} S13, S20, and S21. While this may suggest an intricate order-disorder competition in some cases, the energy differences among the clusters are approximately 1\,meV/atom. With such slight differences, one can anticipate the formation of a homogenous, regular substitutional solid solution with no traces of intergrowth precipitates, ordered intermetallics, or clustering tendencies. This notion is further supported by the fact that the complete randomness is achieved even below 50\,K across the entire solubility range, as the Kullback-Leibler divergence goes to zero very rapidly  (see Supplementary Figure~33). The tendency towards random cluster distribution is also apparent from the symmetric Gibbs mixing free energy parabola around $x=0.5$, with no binodal or spinodal phase decompositions.     

\textbf{Crystal structure}. The expected behavior of the lattice parameters as a function of the composition also indicates a proper solid solution, as depicted in Figure~\ref{fig:nbv_prop}(c). The configurationally averaged lattice constant $a$ fulfills Vegard's law, decreasing linearly with the rise of the calcium content, which is consistent with the similar empirical atomic radii of yttrium and calcium. The lattice mismatch between the end members is only 3.30\,\%. Such a small percentage difference, allied with YH$_6$--CaH$_6$ following the other Hume-Rothery rules, supports the formation of a regular solid solution under normal growth conditions, notwithstanding the inherent challenges associated with synthesizing high-pressure hydrides in diamond anvil cells.   

\textbf{Superconductivity}. Although following Vegard's law and Humme-Rothery's rules, the Y$_{1-x}$Ca$_x$H$_6$ solid solution presents a surprisingly complex behavior regarding the superconducting properties. For estimating the \tc{} for Y$_{1-x}$Ca$_x$H$_6$, we have used the SISSO machine-learned modified Allen-Dynes equation as proposed in Ref.~\cite{xie2022}. The Allen-Dynes equation gives good predictions for low-{\tc} superconductors but underestimates the \tc{} for high-\tc{} superconductors such as hydrides~\cite{lucrezi2024}. Recently, a machine-learned modified Allen-Dynes equation was proposed using symbolic regression and the SISSO framework~\cite{xie2022}:
\begin{align}
    \label{eq:UF}
    T_{\text{c}} = \dfrac{f_{\omega}f_{\mu}\omega_{\text{log}}}{1.20}\exp\left(-\dfrac{1.04(1+\lambda)}{\lambda - \mu^{*}(1 + 0.62\lambda)}\right),
\end{align}
with
\begin{align}
    f_{\omega} = 1.92\left(\dfrac{\lambda + \dfrac{\omega_{\log}}{\overline{\omega}_{2}} - \sqrt[3]{\mu^{*}}}{\sqrt{\lambda}\exp\left(\dfrac{\omega_{\log}}{\overline{\omega}_{2}}\right)}\right) - 0.08,
    f_{\mu} = \dfrac{6.86\exp\left(\dfrac{-\lambda}{\mu^{*}}\right)}{\dfrac{1}{\lambda} - \mu^{*} - \dfrac{\omega_{\log}}{\overline{\omega}_{2}}} + 1,
\end{align}
where $f_{\omega}$ and $f_{\mu}$ are the two machine-learned prefactors. This formula performs equally as good as Allen-Dynes one for low-{\tc} but is more accurate for high-{\tc} superconductors, the reason why it will be used for the superhydrides.

The superconducting \tc{}$(x)$, $\alpha^2F(x,\omega)$, and $D(x)$ for Y$_{1-x}$Ca$_x$H$_6$ at 200\,GPa are shown in Figures~\ref{fig:ycah6_prop}(d)-(g). The plot displays the experimentally obtained \tc{} values for YH$_6$ (\tc{} = 211\,K~\cite{kong2021}) and CaH$_6$ (\tc{} = 204\,K~\cite{ma2022}), which were utilized to determine the \mustar{} in Eq.~(\ref{eq:UF}). The resulting values for the Coulomb pseudopotentials are 0.214 and 0.223 for YH$_6$ and CaH$_6$, respectively. For the off-stoichiometric compounds, we applied the linear interpolation approximation for \mustar{}. 

Interestingly, the variation of \tc{} as a function of composition does not follow a linear trend. In fact, according to the EGQCA, even though pure CaH$_6$ has a lower \tc{} than pure YH$_6$, creating an alloy with 28\,\% calcium content increase \tc{} by 5\,K. Above $x$ = 0.28, the \tc{} decreases linearly until Ca-87~at.\,\%, where the lowest \tc{} of 199\,K in Y$_{1-x}$Ca$_x$H$_6$ is attained. Conversely, as the yttrium content decreases in the CaH$_6$ phase, the \tc{} increases again up to 204\,K.   

We attribute this to the behavior of \af{} as a function of composition, as better visualized in Figure~\ref{fig:ycah6_prop}(f). The \nef{} decreases linearly with higher calcium concentrations, while the \omegalog{} exhibits a parabolic shape, with its minimum around $x=0.3$ (Supplementary Figure~39). However, the electronic low-energy states strongly couple with the high-frequency phonon modes between 100\,meV and 200\,meV in the Y-rich clusters, resulting in an enhanced electron-phonon coupling strength. In the compositional range where we observed the linear-like decrease of \tc{} in Y$_{1-x}$Ca$_x$H$_6$, there is a softening of the Eliashberg spectral function, with a reduced contribution of the high-frequency phonon modes to the electron-phonon coupling. This leads to a sinusoidal-like shape of $\lambda$ as a function of composition, as observed for \tc{} in Figure~\ref{fig:ycah6_prop}(d).        

These characteristics render YH$_6$--CaH$_6$ a promising platform for developing and improving the current techniques and methodologies for designing superhydride alloys. The relative ease of synthesizing Y$_{1-x}$Ca$_x$H$_6$ compounds as regular solid solutions, owing to their favorable thermodynamics and structural properties, provides experimentalists with a significant advantage in improving growth techniques to produce disordered materials and fine-tune compositions under extreme conditions. The predicted superconducting behavior for Y$_{1-x}$Ca$_x$H$_6$ alloys also highlights the potential for complex features in the \tc{} of disordered superhydrides, extending beyond the typical linear or dome-like behaviors observed in superconducting solid solutions. Additionally, from a theoretical perspective, it would be interesting to further incorporate quantum harmonic lattice effects within the EGQCA formalism with the help of moment tensor potentials, as explored in Refs.~\cite{lucrezi2023,lucrezi2024_2} —- a topic we are currently working with.

\section{Summary}
\label{sec:summary}

We have presented an extended formulation of the GQCA model including vibrational effects, and showcased its applicability for modeling alloyed superconductors. EGQCA is a simple yet accurate statistical thermodynamic model that describes the alloy as an ensemble of statistically and energetically independent clusters and provides the probability of occurrence of each cluster as a function of composition and temperature. Throughout this work, we demonstrate the efficiency and robustness of EGQCA in describing and designing superconducting alloys, and what are the best practices for employing this method for complex systems.

As a proof of concept, we applied EGQCA to the Al-doped MgB$_2$, niobium alloyed with titanium and vanadium, and predicted the formation of the high-\tc{} Y$_{1-x}$Ca$_x$H$_6$ solid solution at 200\,GPa. 

EGQCA predictions for Mg$_{1-x}$Al$_x$B$_2$ align excellently with available experimental data. Despite limitations due to the reduced number of cluster configurations, EGQCA provides valuable physical insights and sheds light on outcomes of advanced experimental techniques. The agreement between experimental data and the configurationally averaged lattice parameters and {\tc} of Mg$_{1-x}$Al$_x$B$_2$ is noteworthy. Nearly all experimental data points fall within EGQCA standard deviation. Even without considering anisotropic or anharmonic effects, EGQCA captures all the primary features of the Mg$_{1-x}$Al$_x$B$_2$ solid solution. Notably, our results suggest the feasibility of producing superior Mg$_{1-x}$Al$_x$B$_2$ single crystals and offer a way to reconcile discrepancies among experiments.

For niobium alloys, we have drawn attention to the existing challenges for correctly describing the screened Coulomb interactions as a function of composition. By assuming an increasing Coulomb pseudopotential \mustar{} for Nb--Ti relative to titanium solute, remarkable agreement with experiments is achieved. However, the linear approximation of \mustar{} for the off-stoichiometric compounds fails to replicate the \tc{} trend of Nb$_{1-x}$V$_x$, suggesting the need for additional adjustable parameters. Nonetheless, EGQCA can capture the intricate \tc{} behavior in this system by only fixing three adjustable \mustar{} parameters, suggesting an enhanced screened Coulomb pseudopotential for disordered compounds.

We also demonstrate that the standard GQCA, without the inclusion of vibrational effects, significantly overestimates the critical temperature of the miscibility gap for Nb$_{1-x}$V$_x$. Conversely, EGQCA matches the experimental value of 1078\,K very well. Thus, our results indicate that vibrational entropy is the driving force for forming the miscibility gap of the $bcc$ phase in the Nb--V system, acting to decrease the critical temperature and increase the solubility limit of the $bcc$ structure.

This work also demonstrates that EGQCA is a powerful tool for designing new alloyed superhydrides. We have modeled the Y$_{1-x}$Ca$_x$H$_6$ for the first time, predicting a homogenous, regular substitutional solid solution without intergrowth precipitates, ordered intermetallics, or clustering tendencies, notwithstanding the inherent challenges associated with synthesizing high-pressure hydrides in diamond anvil cells. While following Vegard's law and Hume-Rothery's rules, the Y$_{1-x}$Ca$_x$H$_6$ solid solution exhibits a complex behavior regarding superconducting properties. The variation of \tc{} as a function of composition does not follow a linear trend. Adding calcium marginally elevates \tc{}, reaching 216\,K at Ca-28~at.\,\%, representing a 2.4\,\% increase. Above $x$ = 0.28, \tc{} decreases linearly until Ca-87~at.\,\%, where the lowest \tc{} of 199\,K in Y$_{1-x}$Ca$_x$H$_6$ is attained. Conversely, as yttrium content decreases in the CaH$_6$ phase, \tc{} increases again up to 204\,K. 

YH$_6$--CaH$_6$ offers a promising platform for developing and refining techniques for designing superhydride alloys. Synthesizing Y$_{1-x}$Ca$_x$H$_6$ compounds as regular solid solutions should be relatively straightforward due to favorable thermodynamic and structural properties. This provides experimentalists with an advantage in enhancing growth techniques and fine-tuning compositions under extreme conditions. The predicted superconducting behavior for Y$_{1-x}$Ca$_x$H$_6$ alloys also suggests the potential for complex features in \tc{} of disordered superhydrides, extending beyond the typical linear or dome-like behaviors observed in superconducting solid solutions.

In summary, EGQCA is a new, efficient methodology for modeling superconducting alloys. With advancements in ab initio computational techniques for computing complex variables, such as the static screened Coulomb potential, and integration with machine-learned inter-atomic potentials, the era of fully \emph{in silico} high-throughput design of superconducting alloys is around the corner. We hope our work inspires further experimental and theoretical investigations into well-known superconducting solid solutions. Finally, we want to highlight that EGQCA particularly possesses a promising potential for engineering superhydride alloys, as the correlation effects and electron-electron interactions are usually smaller compared to the strong electron-phonon mass-enhanced parameters.   

\section{Methods}
\label{sec:methods}

First-principles electronic structure calculations were conducted using Density Functional Theory (DFT) \cite{hohenberg1964} within the Kohn-Sham scheme \cite{kohn1965} employing the pseudopotential approach, as implemented in \textsc{Quantum Espresso} \cite{giannozzi2009, giannozzi2017}. Exchange and correlation effects were treated using scalar-relativistic pseudopotentials within the generalized gradient approximation (GGA) according to Perdew-Burke-Ernzerhof (PBE) parametrization \cite{perdew1996}. Projector augmented wave (PAW) pseudopotentials \cite{blochl1994, kresse1999, dal2014} were utilized for Mg$_{1-x}$Al$_x$B$_2$, while optimized norm-conserving Vanderbilt pseudopotentials~\cite{ONCV1,ONCV2} were employed for Nb$_{1-x}$Ti$_x$, Nb$_{1-x}$V$_x$, and Y$_{1-x}$Ca$_x$H$_6$. Self-consistent-field (SCF) calculations with a convergence threshold of $10^{-10}$\,Ry were carried out using Methfessel-Paxton smearing~\cite{MP-smearing}. Lattice parameters and atomic positions were fully relaxed to achieve a ground-state convergence of 10$^{-7}$\,Ry in total energy and 10$^{-6}$ Ry/a$_0$ for forces acting on the nuclei.

The non-equivalent clusters were generated using the \textsc{Supercell} program~\cite{okhotnikov2016}.

The dynamical matrices and the linear variation of the self-consistent potential were calculated within the density-functional perturbation theory~\cite{DFPT} as implemented in \textsc{Quantum Espresso} \cite{giannozzi2009, giannozzi2017}. The threshold for self-consistency is set to 10$^{-14}$ or lower.

Coulomb pseudopotentials were calculated according to the Morel-Anderson approximation at the \textit{GW} level using the \textsc{SternheimerGW} code \cite{lambert2013}.

Further computational details can be found in the Supplementary Table~1.

\section*{Author Contributions}

PNF conceived the project, conducted the DFT and DFPT calculations, analyzed the results, and wrote the manuscript. PNF and LTFE implemented the GQCA and EGQCA models. RL assisted with the DFPT calculations. IG, MM, and LKT contributed to the GQCA discussions. CH and LTF supervised the project. All authors participated in the discussions and reviewed the manuscript.

\section*{Declaration of competing interest}

The authors declare that they have no known competing financial interests or personal relationships that could have appeared to influence the work reported in this paper.

\section*{Acknowledments}

We thank Lilia Boeri and Markus Aichhorn for fruitful discussions and comments on the
manuscript. PNF and LTFE acknowledge the S\~{a}o Paulo Research Foundation (FAPESP) under Grants 2020/08258-0 and 2021/13441-1. RL and CH acknowledge the Austrian Science Fund (FWF) Project No. P 32144-N36. LKT acknowledges the Conselho Nacional de Desenvolvimento Cient\'{\i}fico e Tecnol\'{o}gico (CNPq) under Grant 316081/2023-0. This study was supported by the Coordena\c{c}\~{a}o de Aperfei\c{c}oamento de Pessoal de N\'{\i}vel Superior (CAPES) - Brasil - Finance Code 001, and was conducted using computational resources of the dCluster of the Graz University of Technology, and the VSC5 of the Vienna Science Cluster (VSC).

\section*{Data availability}

All data used to prepare this manuscript are freely available on the Zenodo repository \cite{zenodo}.


\bibliographystyle{apsrev4-2}
\bibliography{refs}

\end{document}